\documentclass[times,twocolumn,review]{elsarticle}

\usepackage{medima}
\usepackage{framed,multirow}
\usepackage{amssymb}
\usepackage{latexsym}
\usepackage{url}
\usepackage{xcolor}
\usepackage{hyperref}
\usepackage{lineno}
\usepackage{graphicx}
\usepackage{tikz}
\usepackage{comment}
\usepackage{amsmath}
\usepackage{array}
\usepackage{sidecap}
\usepackage{threeparttable}
\usepackage{siunitx}
\newcolumntype{P}[1]{>{\centering\arraybackslash}p{#1}}

\DeclareMathOperator{\dist}{dist}

\newcommand{\R}{\mathbb{R}}

\newcommand{\myparagraph}[1]{\smallskip\noindent\textbf{#1}}

\usepackage{float}

\usepackage{booktabs}
\usepackage{multirow}
\newcommand{\ra}[1]{\renewcommand{\arraystretch}{#1}}
\newsavebox\CBox
\def\textBF#1{\sbox\CBox{#1}\resizebox{\wd\CBox}{\ht\CBox}{\textbf{#1}}}

\newcommand{\superemph}[1]{\textbf{#1}}

\usepackage{subcaption}
\definecolor{newcolor}{rgb}{.8,.349,.1}
\journal{Medical Image Analysis}

\begin{document}
\verso{Fan Wang \textit{et~al.}}
\begin{frontmatter}

\title{TopoTxR: A topology-guided deep convolutional network for breast parenchyma learning on DCE-MRIs} 

\author[1]{Fan \snm{Wang}\corref{cor1}}
\author[1]{Zhilin \snm{Zou}\corref{cor1}}
\author[2]{Nicole \snm{Sakla}}
\author[2]{Luke \snm{Partyka}}
\author[2]{Nil \snm{Rawal}}
\author[7]{Gagandeep \snm{Singh}}
\author[4]{Wei \snm{Zhao}}
\author[1]{Haibin \snm{Ling}}
\author[3,6]{Chuan \snm{Huang}}
\author[5]{Prateek \snm{Prasanna}\corref{cor2}}
\author[5]{Chao \snm{Chen}\corref{cor2}}
\cortext[cor1]{Equal contributions}
\cortext[cor2]{Corresponding author}
\cortext[]{E-mail addresses: fanwang1@cs.stonybrook.edu (F. Wang), Prateek.Prasanna@stonybrook.edu (P.Prasanna), chao.chen.1@stonybrook.edu (C.Chen). }

\address[1]{Department of Computer Science, State University of New York at Stony Brook, NY, USA}
\address[2]{Department of Radiology, Newark Beth Israel Medical Center, NJ, USA}
\address[7]{Department of Radiology, Columbia University Irving Medical Center, NY, USA}
\address[3]{Department of Radiology and Imaging Sciences, Emory University School of Medicine, GA, USA}
\address[4]{Department of Radiology, State University of New York at Stony Brook, NY, USA}
\address[5]{Department of Biomedical Informatics, State University of New York at Stony Brook, NY, USA}
\address[6]{Department of Biomedical Engineering, Georgia Institute of Technology and Emory University, Atlanta, GA, USA}



\begin{abstract}
Characterization of breast parenchyma in dynamic contrast-enhanced magnetic resonance imaging (DCE-MRI) is a challenging task owing to the complexity of underlying tissue structures. Existing quantitative approaches, like radiomics and deep learning models, lack explicit quantification of intricate and subtle parenchymal structures, including fibroglandular tissue. 
To address this, we propose a novel topological approach that explicitly extracts multi-scale topological structures to better approximate breast parenchymal structures, and then incorporates these structures into a deep-learning-based prediction model via an attention mechanism. Our topology-informed deep learning model, \emph{TopoTxR}, leverages topology to provide enhanced insights into tissues critical for disease pathophysiology and treatment response. 
We empirically validate \emph{TopoTxR} using the VICTRE phantom breast dataset, showing that the topological structures extracted by our model effectively approximate the breast parenchymal structures. We further demonstrate \emph{TopoTxR}'s efficacy in predicting response to neoadjuvant chemotherapy. Our qualitative and quantitative analyses suggest differential topological behavior of breast tissue in treatment-na\"ive imaging, in patients who respond favorably to therapy as achieving pathological complete response (pCR) versus those who do not. In a comparative analysis with several baselines on the publicly available I-SPY 1 dataset (N=161, including 47 patients with pCR and 114 without) and the Rutgers proprietary dataset (N=120, with 69 patients achieving pCR and 51 not), \emph{TopoTxR} demonstrates a notable improvement, achieving a 2.6\% increase in accuracy and a 4.6\% enhancement in AUC compared to the state-of-the-art method.




\end{abstract}


\end{frontmatter}


\section{Introduction}
\label{sec:intro}

Breast cancer imaging faces a critical challenge in accurately modeling complex breast parenchyma structures, which change dynamically due to factors like angiogenesis, radiation therapy, and chemotherapy. Utilizing 3D breast imaging such as MRI to capture and model these changes can significantly impact diagnosis, prognosis, and treatment planning. Traditional cancer imaging studies have primarily focused on tumor texture and shape, overlooking valuable information in the tumor microenvironment. Evidence suggests that diagnostic and prognostic insights lie in the peritumoral stroma and parenchyma, where phenotypic diversity arises from factors like immune infiltration, vascularity, and tissue composition. Parameters like fibroglandular tissue pattern and parenchymal enhancement also influence breast cancer risk and treatment responses. To tailor diagnosis and treatment strategies to improve patient care, there is a critical need for innovative quantitative methods to comprehensively understand breast cancer biology by exploring the tumor microenvironment and surrounding parenchyma, which can be routinely observed in imaging scans such as breast MRI. 

Various approaches for breast image analysis have been proposed. Radiomic approaches learn diagnostic and prognostic signatures from breast tumor and surrounding peritumoral regions using radiomics features \citep{Saha2018Radio,van2017computational}. These handcrafted features, inspired by human knowledge, attempt to capture different measurements such as tumor/peritumor texture \citep{braman2017intratumoral}, vessel geometry descriptors \citep{braman2022novel}, and other similar characteristics. However, these approaches have two fundamental limitations. First, they usually lack an explicit modeling of the complex structural pattern of peritumoral stroma and parenchyma. Second, these handcrafted features lack sufficient flexibility to model the heterogeneous breast parenchyma and thus cannot provide the desired level of predictive power in practice, despite abundant interpretability.

On the other hand, data-driven approaches, such as 
deep neural networks~\citep{jun1} and convolutional neural networks (CNNs)~\citep{burt2018deep,jarkman2022generalization,tack2018knee,subasi2023breast,su2023bcr,Zhang2018CNNMRI,zhu2017deeply},
have shown great promise in various domains, as they learn feature representations in an end-to-end manner. 
While direct application of CNNs to MRIs seems promising~\citep{liu2020novel, selvikvaag2018overview, lyu2022Patt, Li2021MRIDeep,Mazurowski18MricnnSurvey, ali2020deep,dalmics2017using,ranem2022detecting}, CNN methods take the whole MRI as a direct input; a large portion of the input volume may be biologically irrelevant and even noisy enough to bias the prediction model. Additionally, a 3D CNN possesses millions of parameters and requires a substantial amount of training data. Unfortunately, obtaining such amount of data is often impractical for controlled clinical trials like the I-SPY 1 trial~\citep{newitt2016multi}. Furthermore, CNNs suffer from the limitation of feature interpretability as they lack direct association with the underlying breast tissue structures.

\begin{figure}
\centering
  \begin{subfigure}{0.158\textwidth}
    \includegraphics[width=\linewidth]{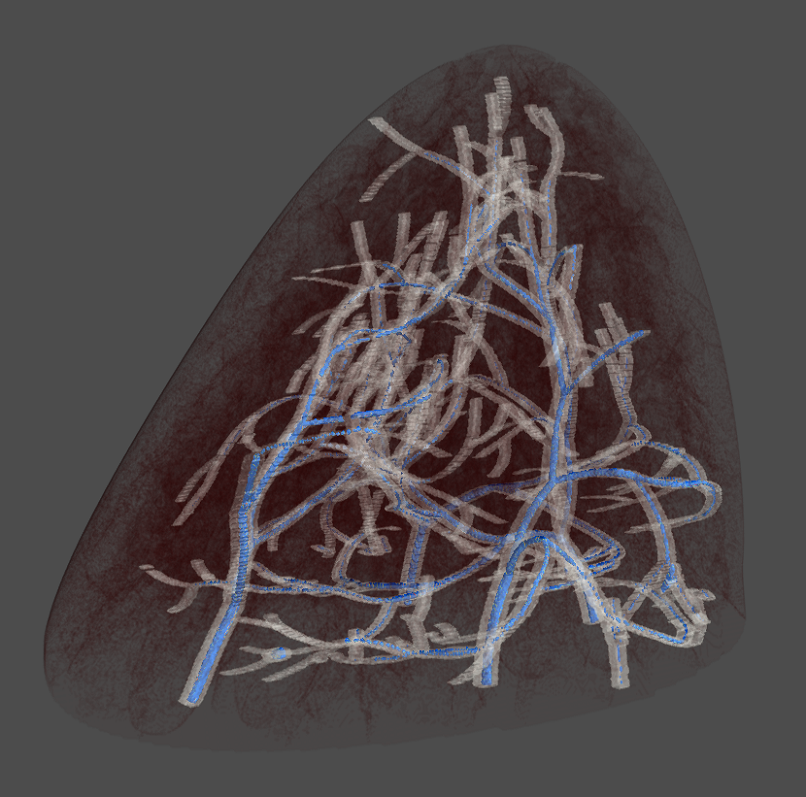}
    \caption{} \label{fig:teaser_victrea}
  \end{subfigure}%
  \hspace*{\fill}   
  \begin{subfigure}{0.158\textwidth}
    \includegraphics[width=\linewidth]{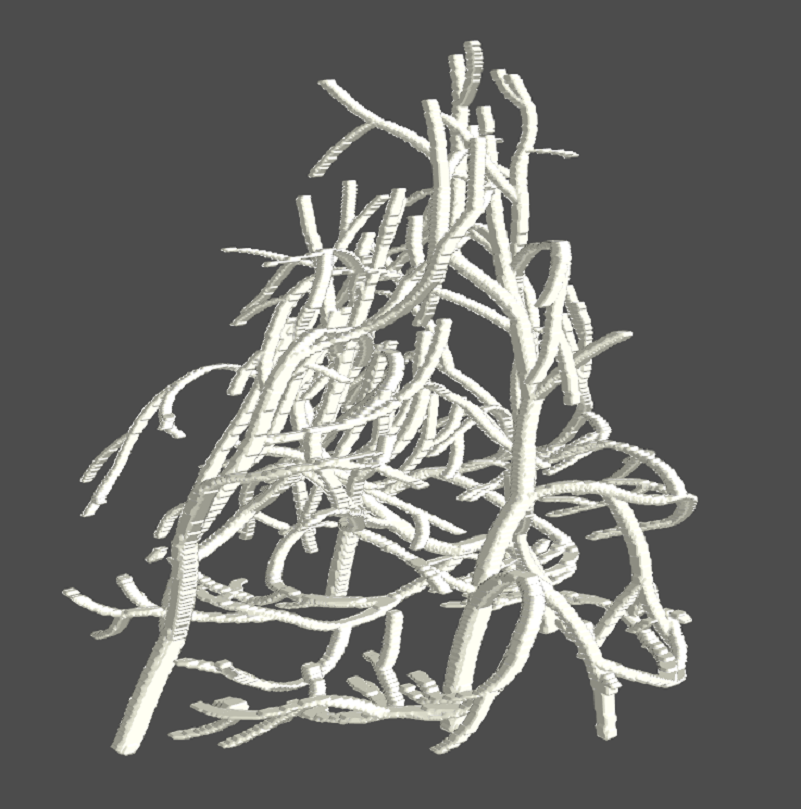}
    \caption{} \label{fig:teaser_victreb}
  \end{subfigure}%
  \hspace*{\fill}   
  \begin{subfigure}{0.158\textwidth}
    \includegraphics[width=\linewidth]{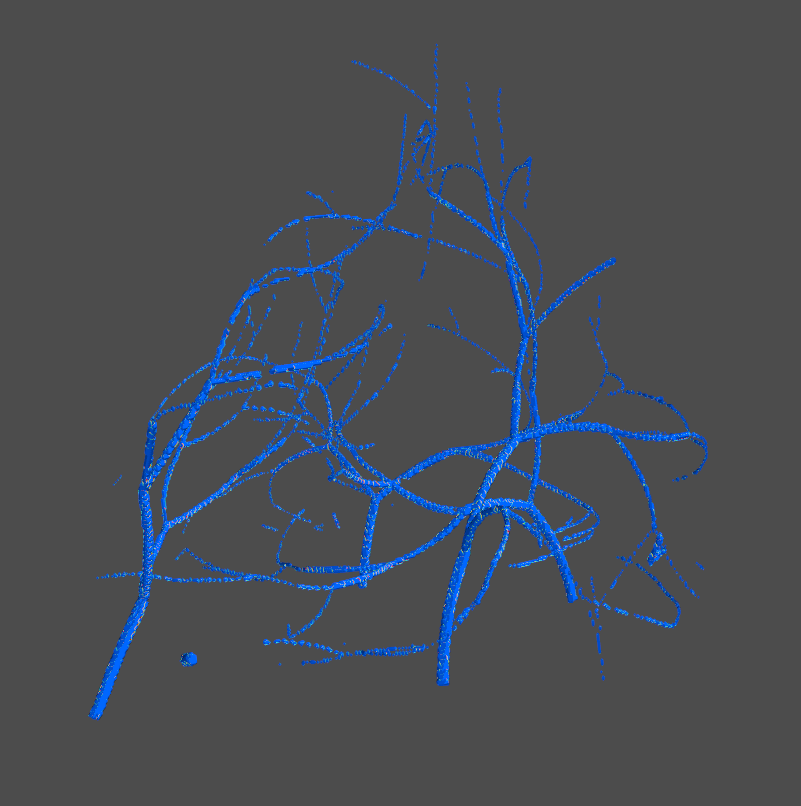}
    \caption{} \label{fig:teaser_victrec}
  \end{subfigure}

\caption{(a): 3D rendering of a phantom breast with highlighted glandular tissues (white) and topological structures (blue); (b): glandular tissues; (c): topological structures.} 
\label{fig:teaser_victre}
\end{figure}

\begin{figure*}[!t]
\centering
\includegraphics[width=.80\textwidth]{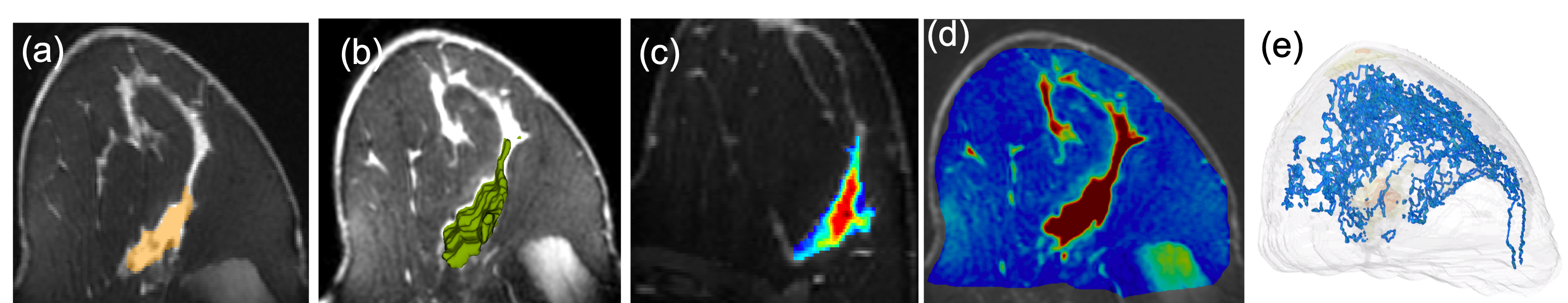}
\caption{ (a) A example MRI image, and different radiomics features such as (b) 3D shape of a tumor, (c) intratumoral texture (Haralick entropy), and (d) whole breast texture (Haralick energy). In (e), we show topological structures from \emph{TopoTxR}, capturing the geometry of fibroglandular tissues.}
\label{fig:features}
\end{figure*}

In this paper, we propose \emph{TopoTxR}, a novel method that overcomes the aforementioned disadvantages. Our method extracts the breast parenchyma structures using the mathematical language of topology. It then effectively incorporates these rich topological structures into deep convolutional neural networks, thereby significantly improving predictive power. By explicitly modeling parenchyma, our approach ensures that predictions are firmly based on biological structures, and thus significantly enhances the performance of the data-intensive CNN model, even with a limited training set.

Our method is based on the theory of persistent homology~\citep{edelsbrunner2010computational}, which extracts 1D (loops) and 2D (bubbles) topological structures with guaranteed robustness~\citep{cohen2005stability}. These structures correspond to curvilinear tissue structures (e.g., ducts, vessels, etc.) and voids enclosed by tissues and glands in their proximity. As shown in Fig.~\ref{fig:features}, compared to previous radiomics features, the topological structures provide a much richer structural context for the modeling of tumor microenvironment. Since these structures are extracted in an unsupervised manner, the quality of their interpretation becomes a key consideration. Using a phantom breast imaging dataset (VICTRE~\citep{VICTRE01}), we validate both quantitatively and qualitatively that these topological structures are reasonable approximations of the breast tissue structures. As illustrated in Fig.~\ref{fig:teaser_victre}, the extracted topological structures delineate the glandular tissues of a phantom breast image. 


 To fully exploit the information carried by these topological structures, we propose a topology-guided deep learning model for breast images. The key idea is to direct the model's attention to voxels adjacent to these topological structures/tissue structures. As the model is focused on a much smaller set of voxels with high biological relevance, it can be efficiently trained with limited MRI data. Meanwhile, the learning outcomes have the potential to connect the biological causes of various breast pathologies with the manifestations observed in the topology of tissue structures. A method closely related to ours, developed by \citep{du2022distilling}, employs a form of weaker topological information known as Betti curves to enhance prediction accuracy. However, the information carried by Betti curves is much more limited when compared to the explicit topological structures used in our approach, leading to suboptimal performance.


While our approach is task-agnostic, we specifically focus on predicting the response to neoadjuvant chemotherapy (NAC) in breast cancer treatment as a practical application. Correct prediction of pathological complete response (pCR) prior to NAC administration can prevent ineffective treatments, reducing unnecessary patient suffering and healthcare costs.
Empirically, we have evaluated our method on the I-SPY 1 dataset~\citep{newitt2016multi} and a proprietary dataset. In these evaluations, \emph{TopoTxR} outperforms various baselines, including radiomics approaches, CNNs trained without topological priors, and other state-of-the-art approaches. This highlights the effectiveness of our topology-centric approach in achieving superior predictive performance.

In summary, we present a novel topological method to characterize parenchyma in breast DCE-MRI, aiming to predict pCR. Our method bridges the two extremes of engineered imaging features and completely data-driven CNNs. Our key contributions are: 
\begin{itemize}
    \item Utilization of persistent homology theory to extract topological structures that closely approximate breast fibroglandular tissue. 
    \item Comprehensive evaluation using a phantom breast imaging dataset to validate the accuracy of the extracted topological structures in approximating breast tissues.
    \item Introduction of a topology-guided spatial attention mechanism designed to direct the focus of 3D CNNs, thereby enhancing their predictive capabilities.
\end{itemize}

This work builds upon our previous conference paper \citep{wang2021topotxr} with four key enhancements: (1) A new topology-guided spatial attention (TGSA) module that explicitly directs the attention of the 3D CNN to biologically relevant sets of voxels with a mask loss. This new module effectively eliminates the need for prior persistent homology computing during the inference stage, making it more practical for real-world applications. (2) An exhaustive empirical evaluation on the VICTRE phantom dataset to validate the accuracy of the topological approximation of breast tissues. (3) Model attention visualization that confirms the proposed model's attention is concentrated on a smaller biologically relevant set of voxels when predicting treatment responses using Gradient-weighted Class Activation Mapping (Grad-CAM) \citep{gradcam}. (4) A new proprietary dataset along with comprehensive evaluations to demonstrate the proposed model's generalization and versatility across different clinical settings. We have made the source code publicly available via this \href{https://github.com/RNZZL/TopoTxR-A-topology-guided-deep-convolutional-network-for-breast-parenchyma-learning-on-DCE-MRIs-2024}{GitHub repository} to facilitate the reproducibility of our research.



\subsection{Related Work}
\label{sec:relatedwork}


Quantitative imaging features have been used in conjunction with machine learning classifiers for the prediction of pCR \citep{radiomic_app3, radiomic_app1}. Radiomics approaches, involving analysis of quantitative attributes of tumor texture and shape, have shown promise in the assessment of treatment response.  In particular, such features capture the appearance of the tumors and, more recently, peritumor regions~\citep{peritumoral1, radiomic1}. Such approaches are often limited by their predefined nature, lack of generalizability, reliance on accurate lesion segmentation, and inability to explain phenotypic differences beyond the peritumoral margin. 
CNNs have been previously applied to breast DCE-MRI for pCR prediction~\citep{CNN1,CNN2,CNN4,CNN5}.  Owing to the sub-optimal performance of image-only models, image-based CNN approaches have been fused with non-imaging clinical variables to bolster prediction~\citep{duanmu2020prediction}.

Moreover, there is extensive literature highlighting the use of CNNs for cancer diagnosis in mammography~\citep{kooi2017large, abdelhafiz2019deep, subasi2023breast}. However, mammograms are only 2D projections of 3D tissue structures. This results in a loss of interpretability of the extracted structures. 2D mammography is limited by the loss of interpretability of extracted structures and 3D mammography using tomography is limited by the inability of the mammogram to provide information regarding background parenchymal enhancement or tumoral/peritumoral enhancement kinetics. Moreover, MRI has been shown to be superior to mammography in determining the extent of breast cancer. In contrast, mammography and ultrasound fail to accurately evaluate tumor size in 8\% of cases post-neoadjuvant chemotherapy, as reported in a retrospective review by \citep{akazawa2006preoperative, keune2010accuracy, londero2004locally}. The purpose of post-neoadjuvant chemotherapy examination is to evaluate the extent of residual disease. However, mammography, ultrasound, and physical exams accurately detect only 13-25\% of pCR cases, as indicated in a study by \citep{herrada1997relative, vinnicombe1996primary, von1999maximized}. MRI and ultrasound each remains superior to mammography with respect to residual tumor detection. Notably, MRI is significantly more effective than mammography in identifying cases of multifocal or multicentric disease, as reported by \citep{londero2004locally}.


Topological information, in particular, persistent homology \citep{edelsbrunner2010computational}, provides a robust way to quantify the topological information in an image. This information, encoded as persistence diagrams or persistence barcodes, has found diverse applications in various image analysis tasks, such as cardiac image analysis \citep{wu2017optimal}, brain network analysis \citep{lee2012persistent, jiachen2024}, and neuron image segmentation \citep{hu2019topology}. In recent years, it has been combined with deep neural networks to enforce topological constraints in image segmentation tasks \citep{clough2020topological, hu2019topology, Shit2020Dice, stucki2023topologically}. Abundant work has been done to learn topology from persistence diagrams, for instance, through vectorization \citep{adams2017persistence}, kernel machines \citep{carriere2017sliced, kusano2016persistence, reininghaus2015stable}, and deep neural networks \citep{hofer2017deep}. Additionally, topology has been formulated as graphs, which are then effectively integrated with Graph Neural Networks (GNN) for applications such as pathology image classification \citep{wang2023ccf} and retinal artery/vein classification \citep{mishra2021Graph}. An even weaker topological information called Betti curve is extracted for learning with breast images \citep{du2022distilling}. However, the actual geometric realization of the topological structures, e.g., cycles and bubbles, has not yet been fully explored. These topological structures capture the geometric details of breast tissues, such as fibroglandular tissues, and can be mapped back to the original breast volumes to provide biologically relevant information for further CNN analysis. For the first time, we propose a deep learning method that leverages the geometry of topological structures as an explicit attention mechanism in this paper. 

\section{Methodology}
\label{sec:method}
\begin{figure*}[!t]
\centering
\includegraphics[width=0.9\textwidth]{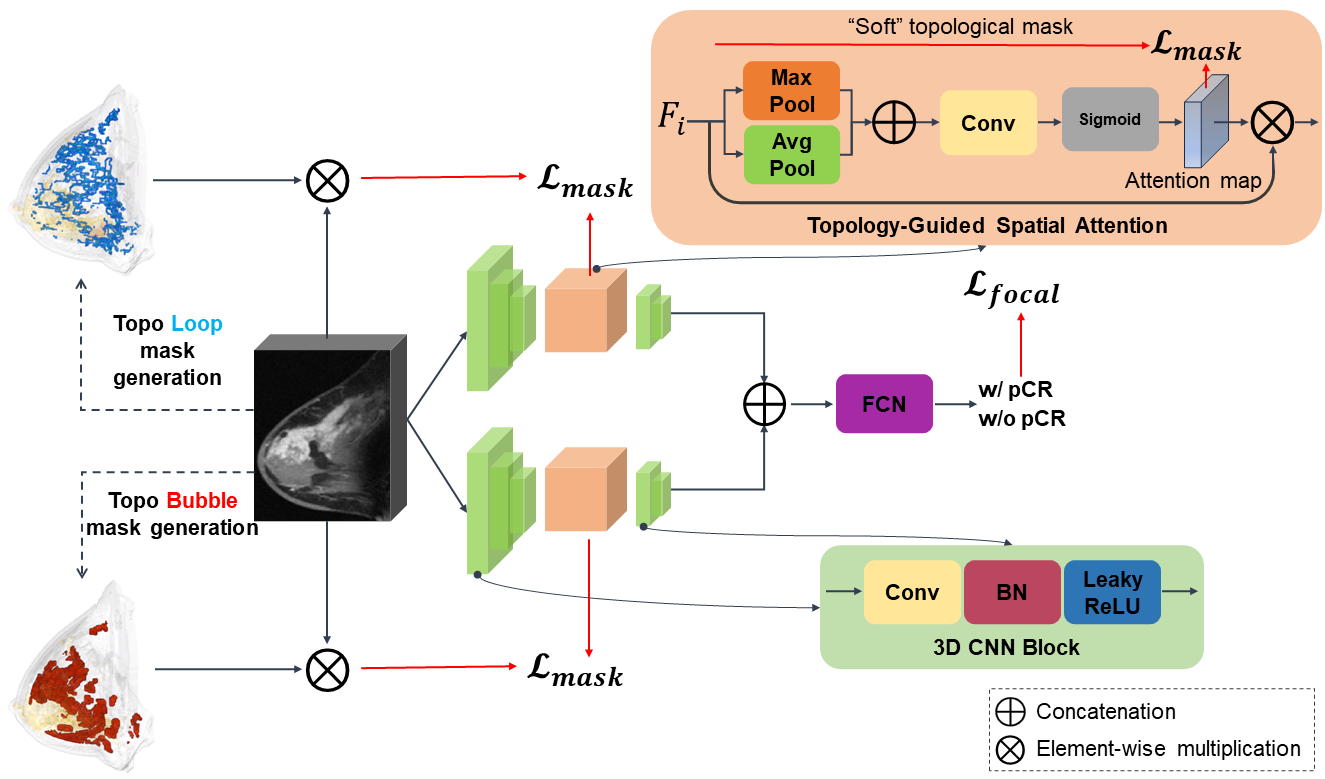}
\caption[overview]{Our proposed TopoTxR pipeline. We extract 1D and 2D topological structures from breast MRI based on persistent homology. Rather than using binary masks, we extract topological structures with intensity values from raw MRIs ("soft" topological masks) for mask loss $\mathcal{L}_{mask}$ supervision. Each 3D CNN branch includes five 3D CNN blocks and a topology-guided spatial attention module (TGSA). The input to TGSA is the feature map from the third convolution layer, $F_i$, while its output to the fourth convolution layer is the generated attention map multiplied by $F_i$. The model features two distinct 3D CNN branches with a fully connected network for pCR prediction.}
\label{fig:overview}
\end{figure*}


We propose a topological method to extract topological structures of high saliency, approximating tissue structures, and utilize these extracted structures as auxiliary information to train a deep convolutional network with raw MRI inputs. Although we focus on training our model for the pCR prediction task, the methodology is versatile enough to be generalized for other tasks. Our approach is detailed in Fig.~\ref{fig:overview}.

We first compute salient topological structures from the input image utilizing persistent homology theory. Topological structures of dimensions 1 and 2, i.e., loops and bubbles, can both correspond to important tissue structures. 1D topological structures capture curvilinear structures such as ducts, vessels, etc. 2D topological structures represent voids enclosed by the tissue structures and their attached glands. These topological structures directly delineate the critical tissue structures with high biological relevance. Thus we hypothesize that by focusing on these tissue structures and their affinities, we can gain pertinent contextual information for pCR prediction.

Subsequently, we introduce a novel 3D CNN framework tailored for breast MRIs that integrates topological structures via an attention mechanism. Our method constructs a custom loss function, combining a mask-guided loss and a refined classification loss, the latter based on focal loss as detailed in Lin et al. ~\citep{lin2017focal}. Notably, we identify two types of pertinent topological structures: loops and bubbles. Our network consists of two separate 3D CNNs, treating the two types of topological structures separately. Empirical evidence demonstrates that both topology types capture complementary structural signatures, proving essential for achieving optimal predictive performance.

Next, we present details of our method, including the background knowledge of persistent homology (Sec.~\ref{sec:method:persistent-homology}), how to compute cycles representing the topological structures (Sec.~\ref{sec:method:cycles}), and our 3D CNN with topology-guided attention (Sec.~\ref{sec:method:topocnn}).

\begin{figure*}[!t]
\centering
\includegraphics[width=.7\textwidth]{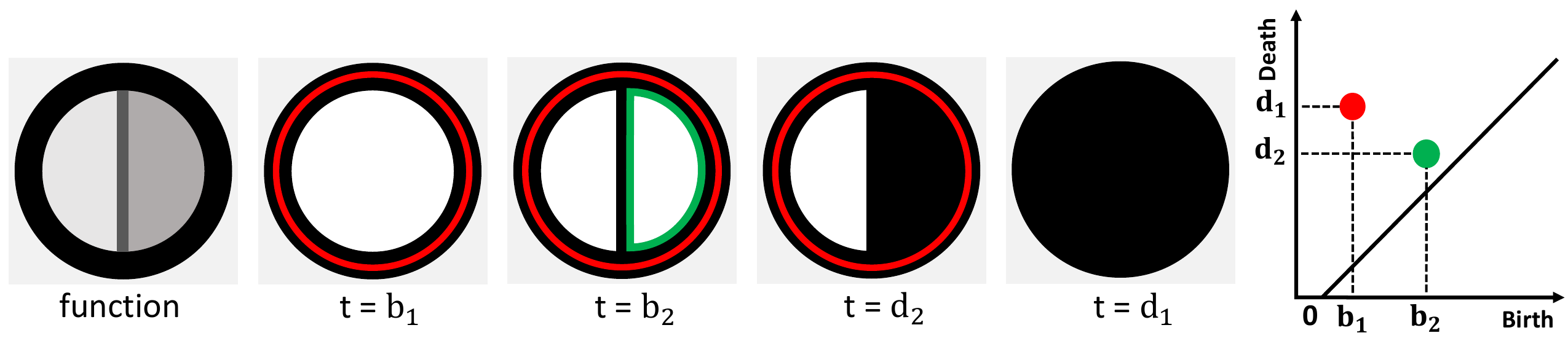}
\caption[ph]{From left to right: a synthetic image $f$, sublevel sets at thresholds $b_{1} < b_{2} < d_{2} < d_{1}$, and the 1D persistence diagram. The red loop represents a 1D structure born at $b_1$ and killed at $d_1$. The green loop represents a 1D structure born at $b_{2}$ and killed at $d_2$. They correspond to the red and green dots respectively in the diagram. }
\label{fig:ph}
\end{figure*}

\subsection{Background: Persistent Homology}
\label{sec:method:persistent-homology}
We review the basics of persistent homology in an intuitive way. Interested readers may refer to \citep{edelsbrunner2010computational} for more details.
Persistent homology extracts the topological information of data observed via a scalar function. Given an image domain, $X$, and a real-valued function $f:X\rightarrow \R$, we can construct a sublevel set $X_{t} =
\{x \in X: f(x) \leq t\}$ where $t$ is a threshold controlling the ``progress"
of sublevel sets. The family of sublevel sets $\mathcal{X} = \{X_{t}\}_{t \in \mathbb{R}}$ defines a filtration, i.e., a family of subsets of $X$ nested with respect to the inclusion: $X_{\alpha} \subseteq X_{\beta}$ if $\alpha \leq \beta$. As the threshold $t$ increases from $-\infty$ to $+\infty$, topological structures such as connected components, handles, and voids appear and disappear. The \emph{birth time} of a topological structure is the threshold $t$ at which the structure appears in the filtration. Similarly, the \emph{death time} is the threshold $t$ at which the structure disappears. Persistent homology tracks the topological changes of sublevel sets $X_{t}$ and encodes them in a \emph{persistence diagram}, i.e., a point set in which each point $(b,d)$ represents a topological structure with birth time $b$ and death time $d$. Its lifespan, i.e., $d-b$, is called the \emph{persistence} of the structure. In practice, we believe persistence is an indicator of the saliency of a topological structure; it has been proven that the persistence of a structure bounds the amount of perturbation one has to inject into the input in order to ``shed off'' the structure \citep{cohen2005stability}. One may compute the number of linearly independent topological structures at different thresholds, $t$. These counts parametrized by $t$ constitute a simpler topological signature called \emph{Betti curves}. We acknowledge that while Betti curves can be directly computed from the persistence diagram, the reverse is not possible.

See Fig.~\ref{fig:ph} for an example function $f$ and its sublevel sets at different thresholds. At time $b_1$, a new handle (delineated by the red cycle $c_{1}$) is created. This handle is later destroyed at time $d_{1}$. Another handle delineated by the green cycle $c_{2}$ is created and killed at $b_{2}$ and $d_{2}$ respectively. The topological changes are summarized in a persistence diagram on the right. Each handle corresponds to a 2D dot in $\mathbb{R}^{2}$, whose $x$ and $y$ coordinates are birth and death times. Their persistence values are $d_1-b_1$ and $d_2-b_2$ respectively.

\begin{figure*}[!t]
\centering
\includegraphics[width=.88\textwidth]{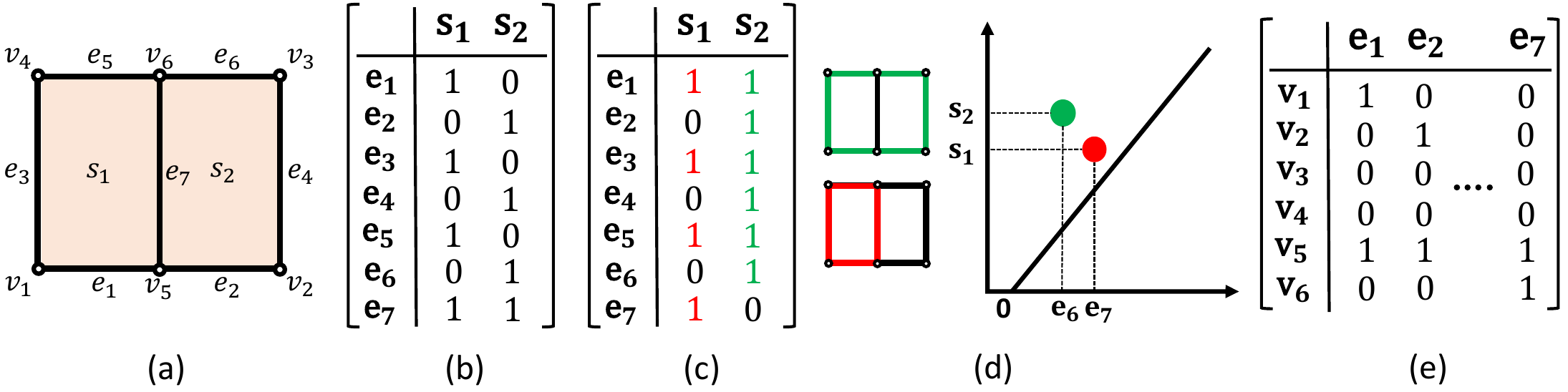}
\caption[cycle]{(a) Example of a cubical complex whose cells are sorted monotonically non-decreasing according to the function values. (b) 2D boundary matrix $\partial$. (c) Reduced boundary matrix. (d) Persistence diagram and resulting topological cycles of $\partial$. (e) 1D boundary matrix.}
\label{fig:cycle}
\end{figure*}

\subsection{Persistence Cycles and Their Computation}
\label{sec:method:cycles}
Although the persistence diagram has been used for topological analysis in various dataset \citep{hu2019topology,wang2020topogan,wu2017optimal}, it only records limited information, i.e., the times at which these topological structures appear/disappear. We hypothesize that a detailed geometric realization of these topological structures can be crucial for learning from images. To this end, we propose extracting these topological structures and integrating them into the learning process. As shown in Fig.~\ref{fig:overview}, we extract loops (blue) to denote 1D topological structures and bubbles (red) for 2D topological structures. These structures are then used to guide the attention mechanism within the neural network.

Next, we formally introduce how different topological structures can be represented with cycles. We also explain how these representative cycles are computed. 
Intuitively, a topological cycle of dimension $p$ is a $p$-manifold without boundary. A 1-dimensional (1D) cycle is a loop. A 2-dimensional (2D) cycle is a bubble. 
A cycle $z$ represents a persistent homology structure if it delineates the structure at its birth. For example, in Fig.~\ref{fig:ph}, the red and the green loops denote the handles born at time $b_1$ and $b_2$, respectively. 

We assume a discretization of the image domain into distinct elements, i.e., vertices (corresponding to voxels), edges connecting adjacent vertices, squares, and cubes. These elements are referred to as 0-, 1-, 2-, and 3-dimensional cells, or simply 0-, 1-, 2-, and 3-cells. A set of $p$-cells constitutes a \emph{$p$-chain}. The \emph{boundary} of a $p$-cell, $\sigma$, is the set of its $(p-1)$-faces. For example, an edge's boundary comprises its two adjacent vertices, a square's boundary includes the four enclosing edges, and a cube's boundary is made up of the six squares surrounding it. The boundary of any $p$-chain, $c$, is the formal sum of the boundaries of all its elements, $\partial(c) = \sum_{\sigma\in c} \partial(\sigma)$, where the sum is the mod-2 sum. In other words, we are taking the exclusive or of the boundaries of all the elements. \footnote{The choice of mod-2 sum is because we focus on homology over $\mathbb{Z}_2$ field, following the common practice in topological data analysis.} For a set of edges forming a path, the boundary consists of the two end vertices. Similarly, for a set of squares forming a patch, its boundary is the loop enclosing the patch, and for a set of cubes, the boundary is the bubble enclosing them.


A $p$-chain is a \emph{$p$-cycle} if its boundary is empty.
All $p$-cycles form the null space of the boundary operator, defined as $\{c : \partial(c)=\emptyset\}$. Every topological structure, formally defined as a \emph{homology class}, can be represented by various cycles, all equivalent under the boundary operator. Formally, given any cycle $z_0$ representing a homology class $h$, we can express $h$ as $\{z_0 + \partial(c) \}$ for all possible chains, $c$, of a higher dimension. We can choose any of the cycles to represent this class. In persistent homology, we can represent each dot in the persistence diagram with one representative cycle at its birth. As depicted in Fig.~\ref{fig:ph}, the red and green cycles represent the two corresponding handles.
Note that the choice of representative cycle is not unique. A relevant question is to choose the shortest representative cycle (i.e., one with the least number of edges) for each dot in the diagram \citep{wu2017optimal,zhang2019heuristic}. In this paper, we focus on choosing a standard representative cycle for computational efficiency.

\myparagraph{Computation of persistent homology and representative cycles.}
We assume a filtration function on a discretization of the image domain. An example discretization of a 2D image is given in Fig.~\ref{fig:cycle}(a). We first sort all cells in increasing order according to their function values. The vertices take the intensity values of their corresponding voxels. For any higher dimensional cell, its function value is the maximum of the values of its vertices. The computation of persistence diagrams is then performed by encoding the $p$-dimensional boundary operator in a binary matrix named \emph{boundary matrices}, $\partial_p$. $\partial_p$ maps $p$-cells to their boundaries. Fig.~\ref{fig:cycle} shows the 1D and 2D boundary matrices of the given complex and its filtration. The 1D boundary matrix is essentially the incidence matrix of the underlying graph (Fig.~\ref{fig:cycle}(e)). 
High dimensional boundary matrices are defined similarly (e.g., a 2D boundary matrix in Fig.~\ref{fig:cycle}(b)). It is essential that the rows and columns of the boundary matrices are arranged so that as we increase the row/column indices, the corresponding cell function values are monotonically non-decreasing.

The persistence diagram is computed by reducing the boundary matrix in a manner akin to Gaussian elimination, but without row or column perturbation. This reduction involves column operations on $\partial$ executed from left to right and from bottom to top. Fig.~\ref{fig:cycle}(c) shows the reduced 2D boundary matrix. Upon reducing the boundary matrices, each non-zero column corresponds to a persistent dot in the persistence diagram, as shown in Fig.\ref{fig:cycle}(d). The reduced column itself is the cycle representing the corresponding topological structure. In this paper, we pay attention to both 1D and 2D cycles, corresponding to loops and bubbles. The extracted cycles will be used to explicitly guide 3D CNNs for analysis.
The computation of representative cycles is of the same complexity as the computation of persistent homology. In theory, it takes $O(n^\omega)$ time ($\omega \approx 2.37$ is the exponent in the matrix multiplication time, i.e., time to multiply two $n\times n$ matrices) \citep{milosavljevic2011zigzag}. Here $n$ is the number of voxels in an image. In practice, computing all cycles for an input image of size ($256^3$) takes approximately 5 minutes.

\subsection{A 3D CNN with Topology-Guided Spatial Attention}
\label{sec:method:topocnn}
To incorporate the generated topological structures into a deep learning classifier, we propose a novel topology-guided spatial attention module. Using a new loss that enforces the spatial attention map to be similar to the topological structures, we direct the model's attention to the vicinity of breast tissues, thus achieving better prediction power.
We first explain the generation of the topological mask, i.e., the union of all topological structures. Next, we explain how to use the topological masks to guide the attention of a deep neural network. 

\myparagraph{Generating topological masks.}
We initiate by extracting topological structures which are represented as sets of voxels. The union of these voxel sets is then used to form a mask that closely approximates biologically pertinent regions.

To ascertain persistence and topological cycles that resemble breast tissue, we invert the MRI image, denoted as \(f = -I\), causing the tissue structures to correspond to regions with lower intensity values. Subsequent to this computation, we pinpoint topological cycles represented by dots on the persistence diagram with high persistence. It is theoretically well understood that dots with lower persistence are less likely to represent genuine signals \citep{cohen2005stability}. As such, we exclusively focus on high-persistence dots, which are indicative of more prominent structures and have a higher likelihood of representing true tissue structures. The persistence threshold is a hyper-parameter adjusted empirically.

Upon selecting topological structures based on persistence, we use the voxels from their representative cycles to craft topological masks. We generate two distinct binary 3D masks that represent the 1D and 2D topological cycles. Empirical observations have revealed that a "soft" mask conveys more detailed information. Rather than directly employing binary masks, the foreground voxels are populated with their inherent image intensity values. Given that all masked MRIs are padded to a size of \(256^3\), we subsequently produce two topological masks, \(M^1_{\text{topo}}\) and \(M^2_{\text{topo}}\), which pertain to the 1D and 2D topological cycles, respectively.

\myparagraph{Topology-guided spatial attention.} In vision-related tasks, spatial attention creates an attention map by harnessing the inter-spatial relationships among features \citep{woo2018cbam}. In the realm of medical tasks, attention mechanisms have been effectively applied to medical image registration \citep{song2022att} and drug response prediction \citep{Feng2021Att}. TopoTxR employs this type of attention to focus on topological structures. Unlike traditional spatial attention, mask-guided attention integrates masks to sharpen model focus which offers a more targeted approach \citep{pang2019}. With these masks defining explicit regions of interest, the model is adept at selectively concentrating on the most pertinent sections, while minimizing attention to regions that might contain noise or are of lesser significance based on mask guidance. This not only streamlines the training process but also bolsters performance. Our work introduces the topology-guided spatial attention module, as illustrated in Fig. \ref{fig:overview}. This module generates a spatial attention map, which is supervised by the aforementioned "soft" topology masks.

We employ distinct 3D CNN networks for both the 1D and 2D branches, each maintaining the same architecture. Every 3D CNN is structured with five 3D convolution layers, with each layer being succeeded by a batch normalization layer and then a LeakyReLU activation. In the topology-guided attention module, the feature maps from the third convolution layer, referred to as \(F_i\) (where \(i\) signifies the branch dimension), are initially processed through both average-pooling and max-pooling operations independently. These pooled results are then concatenated, serving as input for a subsequent convolution layer. The convolution layer's output passes through a sigmoid activation function ,resulting in the generation of attention maps, denoted as \(M_{i}\) where \(i \in [1, 2]\).

To emphasize the regions highlighted by the attention maps, we conduct an element-wise multiplication of \(F_i\) and \(M_{i}\). The product, \(M_{i} \odot F_{i}\), is channeled as input into the fourth layer of the 3D CNN. Outputs from both 3D CNN branches are subsequently vectorized and concatenated. This combined output feeds into the fully convolutional network, culminating in a binary classification. In our ablation study, we further investigate the optimal layer from which to source input for the attention module.

\begin{table}[!t]
\small
\begin{threeparttable}[t]
\centering
\caption{{Average percentage of each tissue type over 25 samples from each phantom breast profile.}} 
\label{table:vicpercentage}
\ra{1.1}
\begin{tabular}{@{}r | P{1.4cm} P{1.4cm} P{1.4cm} P{1.4cm} @{}}
\toprule
\multicolumn{1}{r}{}  
& \multicolumn{1}{c}{\textBF{Dense}} 
& \multicolumn{1}{c}{\textBF{Fatty}} 
& \multicolumn{1}{c}{\textBF{Hetero}} 
& \multicolumn{1}{c}{\textBF{Scattered}}\\
\cmidrule{1-5}
Fat       & 28.59\% & 79.82\% & 49.97\% & 70.44\%\\
Skin      & 4.34\% & 3.09\% & 3.89\% & 3.26\%\\
Glandular & 42.19\% & 6.43\% & 28.45\% & 12.50\%\\
Nipple    & 0.11\% & 0.02\% & 0.07\% & 0.03\%\\
Muscle    & 22.00\% & 8.71\% & 15.18\% & 11.67\%\\\
Ligament  & 1.35\% & 1.68\% & 1.51\% & 1.61\%\\
TDLU\tnote{1}  & 0.31\% & 0.01\% & 0.13\% & 0.03\%\\
Duct      & 0.22\% & 0.07\% & 0.12\% & 0.06\%\\
Artery    & 0.38\% & 0.07\% & 0.30\% & 0.17\%\\
Vein      & 0.49\% & 0.09\% & 0.39\% & 0.21\%\\
\bottomrule
\end{tabular}
\begin{tablenotes}
\item[1] Terminal Duct Lobular Unit
\end{tablenotes}
\end{threeparttable}
\end{table}

The generation of spatial attention masks is guided by "soft" topological masks. This guidance ensures their similarity is captured through a mean squared error (MSE) loss. The loss for our topology-guided spatial attention module, denoted as $\mathcal{L}_{mask}$, is defined as:
\begin{equation}
    \mathcal{L}_{\text{mask}} = \left \|M_{1} - M^1_{\text{topo}} \right \|^2_2 + \left \|M_{2} - M^2_{\text{topo}} \right \|^2_2,
    \label{eq:lmask}
\end{equation}
where 
$M^1_{topo}$ and $M^2_{topo}$ represent the topological masks for 1D and 2D dimensions, respectively.

The focal loss enforces a well-balanced classification performance \citep{lin2017focal}. The focal loss, denoted as $\mathcal{L}_{focal}$, is defined as:
\begin{equation}
    \mathcal{L}_{focal}(p) = -\theta (1-p)^\gamma y_{gt}\log(p)-(1-\theta)p^\gamma(1-y_{gt})\log(1-p)
    \label{eq:lfocal}
\end{equation}
In this context, $y_{gt}$ is the ground truth of the classification, while $p$ indicates the probability of the model prediction. Focal loss incorporates two hyperparameters: $\theta$ and $\gamma$. The parameter $\theta$ is responsible for balancing the weights of positive and negative sample losses, whereas $\theta$ amplifies the loss contribution from hard samples. The mask loss and focal loss together form our overall loss function:
\begin{equation}
    \mathcal{L}_{All} = \mathcal{L}_{focal}+\lambda\mathcal{L}_{mask},
    \label{eq:lall}
\end{equation}
where $\lambda$ is a hyper-parameter.

\section{Topological Approximation of Breast Tissue: Validation on a Phantom Dataset}
\label{sec:validation}
\begin{figure*}[!t]
\centering
\includegraphics[width=.80\textwidth]{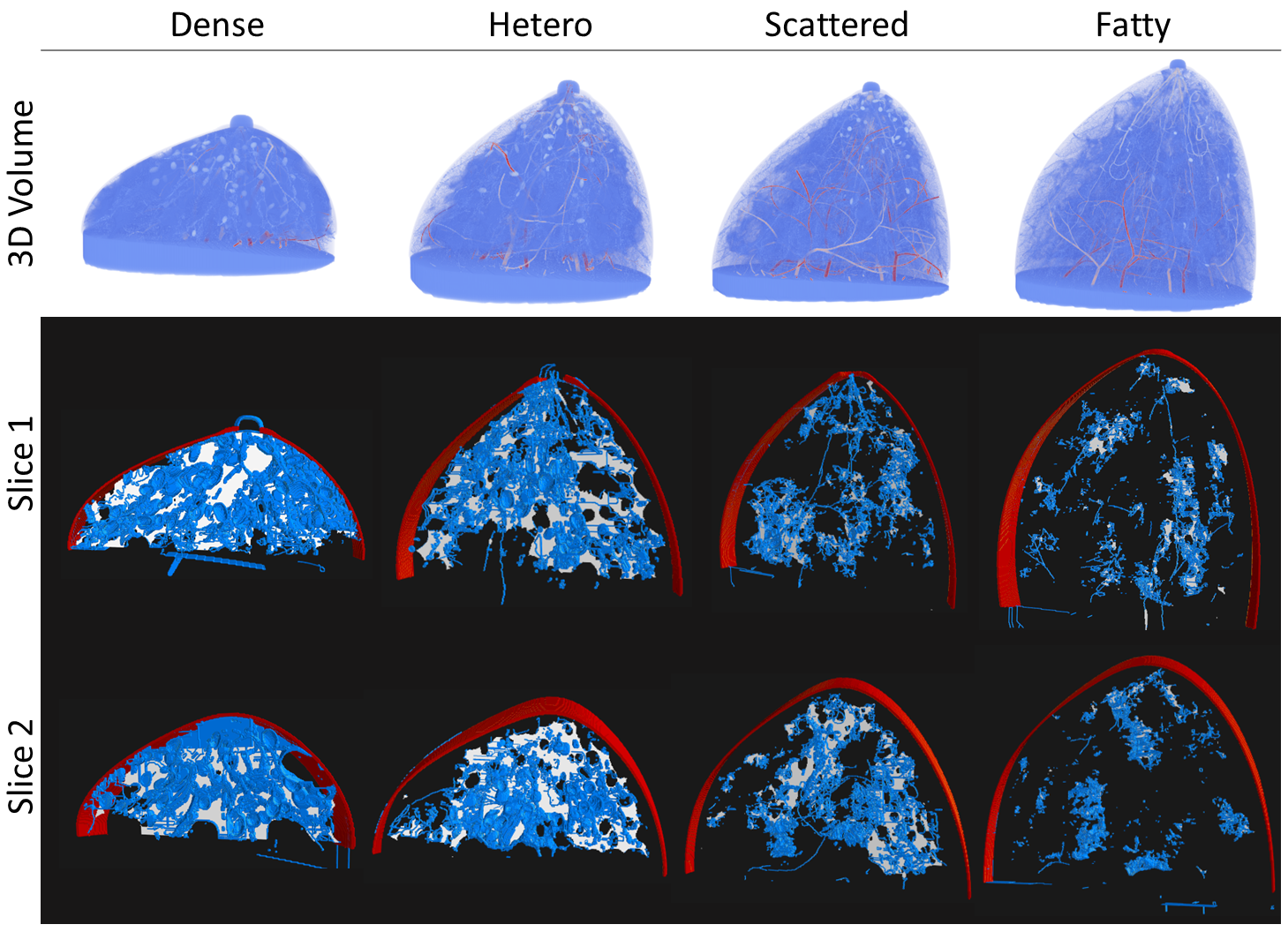}
\caption[ph]{Row 1: 3D renderings of VICTRE phantom breasts of four distinct profiles. Rows 2 and 3: two slices at different positions of the corresponding breast phantoms. Red: 1-voxel width breast outline; blue: extracted topological structures;  white: ground truth breast tissues. Each slice's rendering includes several additional slices around the target cross sections for detailed examination.}
\label{fig:val01}
\end{figure*}

\begin{table}[!t]
\small
\centering
\caption{T1/T2 values and standard deviations for each breast tissue type.} 
\label{table:t1t2}
\ra{1.1}
\begin{tabular}{@{}r | P{2.5cm} P{2.5cm} @{}}
\toprule
\multicolumn{1}{r}{} 
& \multicolumn{1}{c}{\textBF{Mean T1 + Std. Dev.(ms)}} 
& \multicolumn{1}{c}{\textBF{Mean T2 + Std. Dev. (ms)}}\\
\cmidrule{1-3}
Fat            & 366.78  $\pm{7.75}$   & 52.96  $\pm{1.54}$\\
Skin           & 887.00  $\pm{92.00}$  & 22.30  $\pm{7.00}$\\
Glandular      & 1444.83 $\pm{92.70}$  & 54.36  $\pm{9.35}$\\
Nipple         & 796.00  $\pm{21.00}$  & 63.00  $\pm{4.00}$\\
Muscle         & 1232.90 $\pm{255.00}$ & 37.20  $\pm{9.80}$\\
Ligament       & 400.00  $\pm{10.00}$  & 40.00  $\pm{2.00}$\\
TDLU\tnote{1}  & 1444.83 $\pm{92.70}$  & 54.36  $\pm{9.35}$\\
Duct           & 796.00  $\pm{21.00}$  & 63.00  $\pm{4.00}$\\
Artery         & 1984.40 $\pm{146.70}$ & 275.00 $\pm{0.00}$\\
Vein           & 1984.40 $\pm{146.70}$ & 275.00 $\pm{0.00}$\\
\bottomrule
\end{tabular}
\end{table}

The key hypothesis of this paper is that the topological structures, extracted in an unsupervised manner, are good approximations of breast tissue. This section empirically validates this hypothesis.

Ideally, we would like to observe a significant overlap between the true breast tissue and the extracted topological structures. Unfortunately, obtaining ground truth breast tissue from MRIs is very demanding in both time and expertise. Instead, we resort to the VICTRE Breast Phantom data to generate synthetic breast volumes with ground truth breast tissue. This phantom dataset is developed as part of the VICTRE (Virtual Imaging Clinical Trials for Regulatory Evaluation) project by the FDA (Food and Drug Administration) \citep{VICTRE01}. The project produces the VICTRE in-silico trials which provide Monte Carlo simulations of digital mammography (DM) and digital breast tomosynthesis (DBT) exams. Although originally designed for DM and DBT, we can synthesize MRIs using the same phantom data. By comparing with the ground truth tissue, we can validate the approximation quality of the extracted 1D and 2D topological structures.

\subsection{VICTRE synthetic dataset and the synthesized MRIs} 
The VICTRE Breast Phantom generates various types of synthesized breast image volumes based on tunable parameters. Here, we adopt the default configurations used in the VICTRE trial to generate four different profiles of breast phantom densities, covering different real world scenarios: dense, heterogeneously dense, scattered density, and fatty. A synthetic 3D breast phantom is a multi-class mapping in which each voxel is assigned with one of 10 tissue types: fat, skin, glandular, nipple, muscle, ligament, terminal duct lobular unit (TDLU), duct, artery, and vein. Table~\ref{table:vicpercentage} summarizes the composition of tissue types for each profile. Figure \ref{fig:val01} shows sample volumes for each of the four profiles. {In experiments, we generate 25 sample volumes for each profile (100 in total) and simulate their corresponding MRIs for validation.}

Using the 100 synthetic breast volumes, we perform simulations to generate 100 different synthetic T1-weighted MRIs. The T1/T2 weighted values for each tissue type are listed in Table~\ref{table:t1t2}. For each tissue type within a specific breast volume, we sample a single value for both T1 and T2. This ensures that all voxels associated with that particular tissue type have identical values within that breast volume. However, for different breast volumes, we derive a new set of T1 and T2 values for each tissue type based on the known distribution.
The simulated MRI signal is then computed using:
$$
S = k[H]\frac{\sin{\alpha}(1 - e^{-TR/T1})}{(1 - (\cos{\alpha})e^{-TR/T1})}e^{-TE/T2}
$$
where $k[H]$ is the spin density, $\alpha$ stands for the flip angle, and TR (repetition time), and TE (echo time) are MR acquisition parameters. In our experiments, we set $k[H] = 1$, $\alpha = \ang{6}$, $TR = 50$ ms, and $TE = 10$ ms to mimic different T1-weighted images. The simulated MR images are processed by our algorithm to extract topological structures. These extracted structures will then be compared with the ground truth tissues, specifically the synthesized breast image volumes. Figure \ref{fig:val01} shows the corresponding MRI images of the four phantom volumes.


\begin{table*}[!t]
\small
\centering
\caption{{Evaluation of the approximation quality of topological structures using topological precision and recall.}} 
\label{table:vicovp}
\ra{1.1}
\begin{tabular}{P{1.6cm} P{1.6cm} P{1.6cm} P{1.6cm} | P{1.6cm} P{1.6cm} P{1.6cm} P{1.6cm}}
\toprule
\multicolumn{4}{c|}{\textbf{Topological Precision}$\uparrow$} & \multicolumn{4}{c}{\textbf{Topological Recall} $\uparrow$}\\
\cline{1-8}
\multicolumn{1}{c}{\textBF{Dense}} 
& \multicolumn{1}{c}{\textBF{Fatty}} 
& \multicolumn{1}{c}{\textBF{Hetero}}
& \multicolumn{1}{c|}{\textBF{Scattered}} 
& \multicolumn{1}{c}{\textBF{Dense}} 
& \multicolumn{1}{c}{\textBF{Fatty}}
& \multicolumn{1}{c}{\textBF{Hetero}} 
& \multicolumn{1}{c}{\textBF{Scattered}}\\
\hline

$0.93 \pm 0.01$  & $0.95 \pm 0.01$  & $0.97 \pm 0.01$ & $0.96 \pm 0.02$
& $0.26 \pm 0.01$ & $0.20 \pm 0.01$ & $0.18 \pm 0.03$ & $0.17 \pm 0.03$\\
\bottomrule
\end{tabular}
\end{table*}

\begin{table*}[!t]
\small
\centering
\caption{{Evaluation of the approximation quality of topological structures based on the mean distance from topological structures to breast tissues and from breast tissues to topological structures.}}
\label{table:dis}
\ra{1.1}
\begin{tabular}{P{1.6cm} P{1.6cm} P{1.6cm} P{1.6cm} | P{1.6cm} P{1.6cm} P{1.6cm} P{1.6cm}}
\toprule
\multicolumn{4}{c|}{\textbf{Dist.~Tissue to Topo.} $\downarrow$} & \multicolumn{4}{c}{\textbf{Dist.~Topo.~to Tissue} $\downarrow$}\\
\cline{1-8}
\multicolumn{1}{c}{\textBF{Dense}} 
& \multicolumn{1}{c}{\textBF{Fatty}} 
& \multicolumn{1}{c}{\textBF{Hetero}}
& \multicolumn{1}{c|}{\textBF{Scattered}} 
& \multicolumn{1}{c}{\textBF{Dense}} 
& \multicolumn{1}{c}{\textBF{Fatty}}
& \multicolumn{1}{c}{\textBF{Hetero}} 
& \multicolumn{1}{c}{\textBF{Scattered}}\\
\hline
$1.96 \pm 0.08$  & $1.88 \pm 0.06$ & $2.15 \pm 0.09$ & $2.05 \pm 0.07$ &
$1.96 \pm 0.08$  & $15.67 \pm 0.96$ & $5.57 \pm 0.33$ & $11.50 \pm 1.33$\\
\bottomrule
\end{tabular}
\end{table*}



\subsection{Quantitative and qualitative validation} 
We show that the extracted topological structures effectively capture the breast tissues. We apply our method introduced in Section \ref{sec:method:cycles} to the synthetic MRIs to extract topological structures, and then compare these extracted structures with the ground truth tissue in the synthetic volumes. If there is a significant overlap, it indicates that the extracted topological structures closely approximate the actual breast tissue. In our study, breast tissue is defined as the union of all tissue types characterized by rich structures, including glandular tissue, TDLU, ducts, arteries, and veins. Conversely, fat, skin, nipple, muscle, and ligament are categorized separately from breast tissue and are considered as background elements.

The comparison between the topological structures generated by our algorithm and the ground truth breast tissue is detailed in Tables \ref{table:vicovp} and \ref{table:dis}. We denote the predicted topological mask, encompassing the union of 1D and 2D structures, as $M_{topo}$, and the ground truth mask of breast tissue as $M_{tissue}$. To assess this comparison, we employ four different quantitative measures:

\begin{enumerate}
\item \superemph{Topological precision:} the percentage of predicted topological mask being breast tissue, $$\frac{|M_{topo}\cap M_{tissue}|}{|M_{topo}|};$$ 
\item \superemph{Topological recall:} the percentage of breast tissue that is covered by the predicted topological mask,
$$\frac{|M_{topo}\cap M_{tissue}|}{|M_{tissue}|};$$
\item \superemph{Distance from tissue to topological mask:} the mean Euclidean distance from a tissue voxel to its nearest voxel in the topological mask,
$$\frac{1}{|M_{tissue}|}\sum_{x\in M_{tissue}}\min_{y\in M_{topo}} \dist(x,y);$$
\item \superemph{Distance from topological mask to tissue:} the mean Euclidean distance from a voxel in topological mask to its nearest tissue voxel,
$$\frac{1}{|M_{topo}|}\sum_{x\in M_{topo}}\min_{y\in M_{tissue}} \dist(x,y).$$
\end{enumerate}
The topological precision and recall results are presented in Table \ref{table:vicovp}. {Our experiment was conducted on a total of 100 volumes, comprising 25 volumes for each breast profile.} We report the outcomes for these different profiles, providing both the mean and standard deviation. The topological precision numbers indicate that the majority of the extracted topological mask resides within the true breast tissues, showing over 93\% alignment for all four profiles. This high percentage demonstrates the accuracy of the extracted topological mask in delineating the actual tissue. Regarding topological recall, the extracted topological mask's one-voxel thickness means it covers only a portion of the true tissue. In essence, the topological mask acts as a 'skeleton' of the breast tissue. {This observation is highlighted by the topological recall of 26.28\% for dense breast profile. For heterogeneous, scattered, and fatty breast profiles, where the tissue is much sparser, the topological recall shows a lower percentage, ranging between 17\% and 20\%.}

Similar observations are noted in the two distance-based metrics presented in Table \ref{table:dis}. The low mean distance from tissue to topological mask, consistent across all profiles, indicates that the topological mask effectively covers the tissue, with a topological mask voxel usually in close proximity to any tissue voxel. Regarding the distance from topological mask to tissue, we observe low values for dense and heterogeneous profiles, suggesting that the topological mask closely approximates the true tissue in these profiles. Conversely, for scattered and fatty breast profiles, this distance is larger, implying that in these profiles, where the tissue is much sparser, the topological structures tend to be over-predicted, leading to false positive structures that are far from the actual tissue.


\myparagraph{Qualitative validation.} Figure \ref{fig:val01} illustrates how topological structures effectively capture breast tissues. The first row presents 3D renderings of VICTRE-generated synthetic breast phantoms for four distinct profiles. The second and third rows display two slices at different positions of the corresponding breast phantoms, where the topological mask (in blue) and ground truth breast tissues (in white) are shown, alongside the breast outlines (in red). It is important to note that each slice's rendering includes several additional slices around the target cross sections for detailed examination. {It is clear that the blue topological structures closely match the white breast tissues, especially noticeable in the scattered and fatty volumes. This observation highlights that our extracted topological masks effectively mirror the actual breast tissues, confirming the biological significance of these structures.}

\section{Experiments and Results}
\label{sec:experiment}
In this section, we evaluate the efficacy of \emph{TopoTxR} in predicting pCR response to neoadjuvant chemotherapy for breast cancers, utilizing data from both the public ISPY-1 dataset and a proprietary dataset from Rutgers.
Given that \emph{TopoTxR} captures topological structures of high relevance approximating breast parenchymal structures, it offers substantial predictive accuracy in gauging response to neoadjuvant chemotherapy. The structure of this section is as follows: Section ~\ref{data desc} provides details of the ISPY-1 and Rutgers datasets. Section ~\ref{implement} outlines implementation details, including experimental settings and preprocessing steps. Section ~\ref{baselines} provides a detailed discussion of the baseline methods. Subsequently, Section ~\ref{qualres_pcr} presents the quantitative results of our analysis. Finally, Section ~\ref{abltstudy} is dedicated to presenting the findings from our ablation studies.

\subsection{Dataset Description}
\label{data desc}

\myparagraph{I-SPY1:} The I-SPY1 DCE-MRI dataset, available via 'The Cancer Imaging Archive' (TCIA) from the I-SPY1/ACRIN 6657 study~\citep{newitt2016multi}, includes longitudinal breast DCE-MRI scans evaluating the impact of neoadjuvant chemotherapy (NAC) on stage II or III breast cancer patients. All MRIs were performed on a 1.5 Tesla scanner with a breast coil. The study involved patients with T3 tumors, measuring at least 3 cm in diameter, who were undergoing NAC. Initial volume acquisition was pre-contrast, followed by subsequent volume acquisition during and post gadolinium-based contrast agent administration. Pathological complete response (pCR) was used as a measure of the effectiveness of NAC. Our method was validated by predicting pCR using the initial MRIs taken four weeks prior to treatment from the I-SPY1 post-contrast DCE-MRI data. Our analysis included 161 patients, with 47 achieving pCR (average age = 48.8) and 114 not achieving pCR (average age = 48.5).

\myparagraph{Rutgers proprietary dataset:} We also applied our methods to a proprietary dataset comprising 120 patients, among whom 69 achieved pCR and 51 did not. In this study, the first phase T1-weighted fat-suppressed post-contrast DICOM images for each patient, captured 90 seconds post-contrast, were used. All imaging was performed on 1.5T MRI scanners, using gadolinium contrast agents (Gd-DTPA) at a dose of 0.2 mmol/kg, administered at a rate of 2.0 mL/s, followed by a 10 mL saline flush of $0.9\%$. This retrospective case-control study analyzed hospital records spanning from January 1, 2008, to June 30, 2023, and was conducted without direct contact with the participants.

We included 120 women from Robert Wood Johnson Barnabas Health hospitals, specifically Newark Beth Israel Medical Center and Cooperman Barnabas Medical Center, all of whom had confirmed diagnoses of invasive breast carcinoma. Tumors had to be clinically and radiologically measurable post-biopsy, at least 10 mm in diameter. Excluded were patients with prior cytotoxic regimens in the breast of interest, but those treated for contralateral breast cancer were included. Participants were above 18 years of age, neither pregnant nor lactating, had undergone a primary mass core biopsy, and had MRI-compatible health conditions. Tumors needed to be stage I-III, T4, any N, M0, inclusive of clinical or pathological inflammatory cancers, as well as regional stage IV, provided the only site of metastasis was the supraclavicular lymph nodes. Exclusion criteria included recent use (within 30 days) of investigational agents and allergies to compounds resembling MRI contrast dye, preventing contrast-enhanced examinations.

{We would like to highlight that while our model is contingent on high quality dynamic contrast enhanced MRI imaging, its utility across the majority of breast centers is maintained given that most breast practices undergo American College of Radiology (ACR) accreditation. We agree that our model may not be generalizable for those practices which do not use high quality DCE-MRI or ACR accreditation standards. However, to ensure broad applicability, we have also applied our model to the national dataset used in the I-SPY1 trial.}

\subsection{Experiment Details}
\label{implement}
We implemented our model using Pytorch \citep{NEURIPS2019_9015} and conducted all experiments on a single NVIDIA RTX A6000 GPU. Hyperparameters, including learning rate, momentum, loss weight factor, and dropout rate for the dropout layer, were fine-tuned using a grid search. {Specifically, we conducted a 3-fold cross-validation on the I-SPY 1 dataset, testing each combination of parameters in the grid.}
The model training was executed using the Adam optimizer \citep{kingma2014adam}, with the learning rate set at 0.01 and momentum at 0.9. The dropout rate in the last layer was fixed at 0.2, and the weight of the topology-guided mask loss, $\lambda$, was set at 0.01. For the focal loss, the class weight factors, $\theta$ and $\gamma$, were adjusted to 0.68 and 2.0, respectively, in line with the ratio of positive to negative samples in our training set. All input MRIs were of size $256^3$, and the model was trained for 200 epochs with a batch size of 2, amounting to a total training time of approximately four hours. Inference for a single MRI takes less than one minute.

\myparagraph{Preprocessing}. {Preprocessing comprises two phases to prepare the data from both the I-SPY1 and Rutgers proprietary datasets for effective network training. Initially, we focus on noise elimination and standardization of the samples into a uniform shape. Specifically, images with excessive background noise were excluded. For the remaining data, we employed dilation and erosion techniques to further reduce any residual background noise. Subsequently, each MRI volume was cropped to the foreground, ensuring a 2-voxel width margin in each dimension. If any dimension of the cropped MRI volume exceeded 256, we resized it while preserving the original aspect ratio, ensuring that the largest dimension did not exceed 256. To center the foreground in each volume, we applied appropriate padding, resulting in a standardized volume size of \(256 \times 256 \times 256\). This uniformity facilitates seamless integration with deep neural network training. We normalized the volumes' intensity and inverted the intensity values by multiplying each voxel by -1, thus preparing them for subsequent persistent homology computation. The second step involves applying a persistence threshold, selecting dimensions of topological structures, and applying dilation to these structures before using them as inputs for the network training. An ablation study of this second step is detailed in Section~\ref{abltstudy}.}

\begin{table*}[!t]
\begin{minipage}[t]{1\linewidth}
  \centering
  \small
  \caption{Comparative analysis of our proposed method, \emph{TopoTxR}, with baseline methods across four metrics -- accuracy, AUC, specificity, and sensitivity -- on the I-SPY1 dataset, utilizing 10-fold cross-validation. The performance of \emph{TopoTxR} integrated with non-imaging clinical features (denoted as TopoTxR+Clinical) is displayed in the final row of the table. {Models marked with an asterisk (*) represent the 3D versions we developed specifically for this analysis.}}

  \vspace{-2mm}
  \label{table:mainres}
  \begin{tabular}{lccccc}
  \toprule
          & \textbf{Accuracy}$\uparrow$ & \textbf{AUC}$\uparrow$ & \textbf{Specificity}$\uparrow$ & \textbf{Sensitivity}$\uparrow$ \\
    \midrule
{Radiomics}&0.563$\pm{0.085}$ &0.593$\pm{0.098}$ &0.552$\pm{0.180}$  &0.575$\pm{0.081}$  \\
{PD}&0.549$\pm{0.081}$ &0.567$\pm{0.097}$ &0.551$\pm{0.167}$  &0.547$\pm{0.071}$  \\
{Radiomics+PD}&0.563$\pm{0.093}$ &0.587$\pm{0.099}$ &0.592$\pm{0.178}$  &0.534$\pm{0.087}$  \\

{ViT*}&0.770$\pm{0.035}$ &0.676$\pm{0.060}$ &0.508$\pm{0.020}$  &0.874$\pm{0.087}$  \\

{ConvNext*}&0.745$\pm{0.047}$ &0.688$\pm{0.069}$ &0.572$\pm{0.155}$  &0.805$\pm{0.070}$  \\

{ResNext*}&0.740$\pm{0.036}$ &	0.643$\pm{0.088}$ &0.373$\pm{0.229}$  &0.842$\pm{0.108}$  \\

    {DenseNet, \citep{huang2017densely}}&0.814$\pm{0.057}$ &0.821$\pm{0.031}$ &0.827$\pm{0.023}$  &0.816$\pm{0.079}$  \\
    
    {DenseNet* w/ focal}&0.789$\pm{0.043}$ &0.729$\pm{0.053}$ &0.627$\pm{0.136}$  &0.830$\pm{0.071}$  \\
    
    {DenseNet-KD, \citep{du2022distilling}} & 0.905$\pm{0.021}$ & 0.874$\pm{0.016}$ & 0.825$\pm{0.043}$ & 0.923$\pm{0.030}$ \\
    
    \midrule
    {TopoTxR w/o focal or TGSA} & 0.856$\pm{0.046}$ & 0.807$\pm{0.046}$ & 0.664$\pm{0.123}$ & 0.949$\pm{0.056}$ \\
    
    {TopoTxR w/o focal} & 0.913$\pm{0.031}$ & 0.918$\pm{0.031}$ &\textbf{0.944}$\pm{\textbf{0.057}}$ & 0.892$\pm{0.051}$ \\
    
    {Self-Attention w/ focal} & 0.920$\pm{0.030}$ & 0.922$\pm{0.030}$ &0.916$\pm{0.065}$ & 0.927$\pm{0.051}$ \\
    
    \textbf{TopoTxR(Ours)} & \textbf{0.931}$\pm{\textbf{0.027}}$ & 0.917$\pm{0.036}$ & 0.885$\pm{0.068}$ & \textbf{0.950}$\pm{\textbf{0.033}}$ \\
    
    TopoTxR+Clinical & 0.925$\pm{0.023}$ & \textbf{0.926}$\pm{\textbf{0.028}}$ & 0.911$\pm{0.056}$ & 0.940$\pm{0.035}$ \\
    \bottomrule
    \end{tabular}%
\end{minipage}%

\vspace{1em}

\begin{minipage}[t]{1\linewidth}
  \centering
  \small
  \caption{Comparative analysis of our proposed method, \emph{TopoTxR}, with baseline methods across four metrics -- accuracy, AUC, specificity, and sensitivity -- on the proprietary Rutgers dataset, utilizing 10-fold cross-validation. {Models marked with an asterisk (*) represent the 3D versions we developed specifically for this analysis.}}
  \vspace{-2mm}
  \label{table:privateres}
  \begin{tabular}{lccccc}
  \toprule
          & \textbf{Accuracy}$\uparrow$ & \textbf{AUC}$\uparrow$ & \textbf{Specificity}$\uparrow$ & \textbf{Sensitivity}$\uparrow$ \\
    \midrule
{DenseNet*}&0.783$\pm{0.047}$ &0.740$\pm{0.047}$ &0.781 $\pm{0.096}$  &0.698$\pm{0.151}$  \\

{ViT*}&0.800$\pm{0.041}$ &0.757$\pm{0.058}$ &0.819$\pm{0.094}$  &0.696$\pm{0.132}$  \\

{ConvNext*}&0.725$\pm{0.066}$ &0.719$\pm{0.075}$ &0.784$\pm{0.097}$  &0.655$\pm{0.154}$  \\

{ResNext*}&0.758$\pm{0.049}$ &0.745$\pm{0.046}$ &0.795$\pm{0.132}$  &0.695$\pm{0.138}$  \\

{TopoTxR w/o focal or TGSA} & 0.883$\pm{0.041}$ & 0.862$\pm{0.051}$ & \textbf{1.000}$\pm{\textbf{0.000}}$ & 0.724$\pm{0.102}$ \\
    
{TopoTxR w/o TGSA} & 0.900$\pm{0.031}$ & 0.878$\pm{0.042}$ &\textbf{0.988}$\pm{\textbf{0.023}}$ & 0.769$\pm{0.093}$\\
    
\textbf{TopoTxR (Ours)} & \textbf{0.925}$\pm{\textbf{0.028}}$ & \textbf{0.900}$\pm{\textbf{0.055}}$ & 0.934$\pm{0.051}$ & \textbf{0.867}$\pm{\textbf{0.127}}$&\\
\bottomrule
\end{tabular}%
\end{minipage}
\end{table*}%

\subsection{Baselines}
\label{baselines}
We conduct a comparative analysis using various baseline methods on both the I-SPY1 and Rutgers proprietary datasets. Due to significant differences between the Rutgers dataset and I-SPY1, we evaluate them separately. \footnote{The I-SPY1 and Rutgers datasets underwent preprocessing protocols with different hyperparameters, including the selection of minimum tumor size and application of distinct contrast enhancements. Furthermore, the inclusion of patients with stage I or T4 in the Rutgers dataset contributes to variations in the final image formation compared to I-SPY1.} We perform a 10-fold cross-validation on the I-SPY1 dataset and the Rutgers dataset. The baselines are listed below:
\begin{itemize}
  \item \textbf{Radiomics:} We compute a 92-dimensional radiomic signature~\citep{pyradiomics} and train a classifier on this signature. Features are extracted solely from the tumor region.
  \item \textbf{Persistent Diagram (PD):} A classifier is trained using topological features derived from the persistence diagrams (PDs) of the MRI images. While various classifier options are available and behave similarly, we use the sliced Wasserstein kernel distance for PDs as the feature vector \citep{carriere2017sliced}.
  \item \textbf{Radiomics+PD:} This method involves a combination of both radiomic and PD features for classifier training.
\end{itemize}

We implement feature selection for all the aforementioned methods using Mutual Information Difference (MID) and Mutual Information Quotient (MIQ). For all baseline features, an exhaustive search is conducted across all combinations of feature selection schemes and a range of classifiers, including Random Forests, Linear Discriminant Analysis, Quadratic Discriminant Analysis, and Support Vector Machine. We report the results based on the best-performing combinations.

\begin{itemize}
  \item \textbf{DenseNet, {DenseNet with focal loss}, DenseNet-KD:} We perform a comparative analysis of our method against DenseNet, DenseNet with focal loss, and the state-of-the-art method, DenseNet-KD, using the I-SPY1 dataset. DenseNet \citep{huang2017densely} processes raw 3D MRI images directly as input. In contrast, DenseNet-KD \citep{du2022distilling} not only uses 3D MRI images but also integrates topological pseudo labels, particularly Betti number curves, for joint training with the DenseNet backbone. This approach leads to noticeable improvements. The results of this comparative analysis are detailed in Table \ref{table:mainres}.

  \item \textbf{{ConvNext, ResNext, ViT}:}
  {We compare our method against state-of-the-art architectures—ConvNext \citep{liu2022convnet}, ResNext \citep{xie2017aggregated}, and transformer-based ViT \citep{dosovitskiy2020image}—utilizing both the I-SPY 1 dataset and our proprietary Rutgers dataset. The results are presented in Table \ref{table:mainres} for the I-SPY 1 dataset and Table \ref{table:privateres} for the Rutgers dataset. It is important to note that, like DenseNet, ConvNext, ResNext, and ViT were originally designed for 2D inputs. We have developed 3D versions of these models to conduct a comprehensive comparative analysis.}
  
  \item \textbf{TopoTxR+Clinical:} Inspired by the work of Duanmu et al. \citep{duanmu2020prediction}, we gather non-imaging clinical data for the I-SPY1 dataset, which includes demographic information (age, race), Estrogen Receptor Status (ER), Progesterone Receptor Status (PR), Hormone Receptor Status (HR), Human Epidermal Growth Factor Receptor 2 (HER2) Status, and a 3-level HR/HER2 categorization. We utilize this information to construct a non-imaging clinical feature vector. This vector is then concatenated with the feature maps from both 3D CNN branches, forming the input for the fully connected (classification) layer to facilitate pCR prediction.
\end{itemize}

\subsection{Quantitative results}
\label{qualres_pcr}
The quantitative results on the I-SPY1 dataset are presented in Table \ref{table:mainres}. Our method, which incorporates focal loss and topology-guided spatial attention, is compared against various baselines across 4 metrics, namely, accuracy, area under the curve (AUC), specificity, and sensitivity. Notably, our proposed method outperforms the state-of-the-art DenseNet-KD method in all four metrics. This achievement highlights the superiority of the topological mask over the Betti curve, which is the main feature used by DenseNet-KD. The richer information content of the topological mask compared to the Betti curve likely contributes to this enhanced performance. Additionally, our method demonstrates a more balanced specificity and sensitivity, attributed to the effective use of focal loss.

We substitute the topology-guided spatial attention module in our framework with a self-attention module and compare against it. As shown in row 12 of Table \ref{table:mainres}, the topology-guided spatial attention proves to be significantly more effective for pCR prediction. This effectiveness can be attributed to the topological mask's ability to accurately capture biologically relevant breast tissue. Additionally, the results of incorporating non-imaging clinical features into our framework are presented in the last row of Table \ref{table:mainres}. These results indicate that the inclusion of clinical data does not significantly impact the efficacy of our model.

Given the distinct imaging characteristics between the I-SPY1 and Rutgers datasets (as detailed in Section \ref{data desc}), it is not feasible to train \emph{TopoTxR} on one dataset and test it on the other. Consequently, we conduct an independent evaluation using the Rutgers dataset, with the results presented in Table \ref{table:privateres}. {The results presented in both tables demonstrate that our proposed method consistently outperforms the baselines, including state-of-the-art models such as DenseNet, DenseNet-KD, ConvNext, ResNext, and ViT across all datasets. This underscores the effectiveness of our approach and its adept integration of topological knowledge for predicting pCR.}

\myparagraph{Performance analysis across breast tissue density groups.} {We evaluated the performance of the proposed model across various breast tissue densities to assess its clinical applicability. Due to the lack of density labels in the public I-SPY 1 dataset, we classified the samples from our proprietary Rutgers dataset into four groups based on the density of fibroglandular tissue (FGT), as assessed by our radiologists (N.S. and L.P.). The categorization is as follows:}
{
\begin{itemize}
    \item \textbf{Dense:} $>75\%$ of the breast comprises FGT.
    \item \textbf{Heterogeneous:} $51\%-75\%$ of the breast comprises FGT.
    \item \textbf{Scattered:} $25\%-50\%$ of the breast comprises FGT.
    \item \textbf{Fatty:} $<25\%$ of the breast comprises FGT.
\end{itemize}
}

{We now present the performance of our model across the various breast tissue density groups. According to Table \ref{table:density}, our method consistently delivers superior performance across all density groups, achieving perfect results on the fatty volumes. This highlights the robustness of our proposed model across different breast tissue types.}

\begin{table}[!t]
\small
\centering
\caption{{TopoTxR pCR prediction results for four different densities of fibroglandular tissue in the Rutgers dataset.}}
\label{table:density}
\ra{1.1}
\begin{tabular}{@{}r | P{1.2cm} P{1.2cm} P{1.2cm} P{1.2cm} @{}}
\toprule
\multicolumn{1}{r}{}  
& \multicolumn{1}{c}{\textBF{Accuracy}$\uparrow$} 
& \multicolumn{1}{c}{\textBF{AUC}$\uparrow$} 
& \multicolumn{1}{c}{\textBF{Specificity}$\uparrow$} 
& \multicolumn{1}{c}{\textBF{Sensitivity}$\uparrow$}\\
\cmidrule{1-5}
Dense & 0.952 & 0.929 & 0.857 & 1.000 \\ 
Hetero & 0.913 & 0.912& 0.905 & 0.920\\ 
Scattered & 0.902 & 0.898& 0.875 & 0.920 \\ 
Fatty & 1.000 & 1.000& 1.000 & 1.000 \\ 
\bottomrule
\end{tabular}
\end{table}


\begin{table*}[!t]
\small
\centering
\caption{Ablation study results. All numbers are reported based on a 10-fold cross-validation on the I-SPY1 dataset.}
\label{table:abl}
\ra{1.1}
\begin{tabular}{@{}r | P{2cm} P{2cm} P{2cm} P{2cm} @{}}
\toprule
\multicolumn{1}{r}{} 
& \multicolumn{1}{c}{\textBF{Accuracy}$\uparrow$} 
& \multicolumn{1}{c}{\textBF{AUC}$\uparrow$} 
& \multicolumn{1}{c}{\textBF{Specificity}$\uparrow$} 
& \multicolumn{1}{c}{\textBF{Sensitivity}$\uparrow$}\\
\cmidrule{2-5}
\multicolumn{1}{r}{} & \multicolumn{4}{c}{Persistence Threshold}\\ \midrule
$90\%$ Remain& 0.913$\pm{0.026}$ & 0.873$\pm{0.058}$ & 0.817$\pm{0.113}$ & 0.928$\pm{0.052}$\\
$60\%$ Remain& 0.906$\pm{0.036}$ & 0.864$\pm{0.058}$ & 0.769$\pm{0.128}$ & 0.960$\pm{0.039}$\\

\midrule
\multicolumn{1}{r}{} & \multicolumn{4}{c}{Dimension}\\ \midrule
Dimension 1 & 0.900$\pm{0.026}$ & 0.896$\pm{0.035}$ & 0.880$\pm{0.070}$ & 0.912$\pm{0.036}$\\
Dimension 2 & 0.863$\pm{0.038}$ & 0.845$\pm{0.060}$ & 0.792$\pm{0.141}$ & 0.899$\pm{0.055}$\\
{MRI+Dim1+Dim2} & 0.863$\pm{0.042}$ & 0.825$\pm{0.058}$ & 0.675$\pm{0.124}$ & 0.954$\pm{0.043}$\\
{MRI} & 0.633$\pm{0.200}$ & 0.621$\pm{0.102}$ & 0.570$\pm{0.322}$ & 0.673$\pm{0.354}$\\

\midrule
\multicolumn{1}{r}{} & \multicolumn{4}{c}{Dilation Radius}\\ \midrule
Radius 2 & 0.894$\pm{0.043}$ & 0.873$\pm{0.058}$ & 0.817$\pm{0.113}$ & 0.928$\pm{0.052}$\\
Radius 4 & 0.863$\pm{0.034}$ & 0.860$\pm{0.063}$ & 0.827$\pm{0.152}$ & 0.893$\pm{0.045}$\\
Radius 8 & 0.913$\pm{0.031}$ & 0.913$\pm{0.031}$ & 0.911$\pm{0.075}$ & 0.907$\pm{0.061}$\\

\midrule
\multicolumn{1}{r}{} & \multicolumn{4}{c}{Insertion Point for the Topology-Guided Spatial Attention Module}\\ \midrule
After 2\textsuperscript{nd} Conv &0.894$\pm{0.018}$&0.908$\pm{0.025}$&0.919$\pm{0.064}$&0.898$\pm{0.042}$ \\
After 4\textsuperscript{th} Conv &0.889$\pm{0.042}$&0.879$\pm{0.050}$&0.882$\pm{0.104}$&0.876$\pm{0.085}$  \\

\bottomrule
\end{tabular}
\end{table*}

\subsection{Ablation Studies}
\label{abltstudy}
\myparagraph{TopoTxR w/o focal loss or TGSA.} To highlight the improvements introduced by the focal loss and the topology-guided spatial attention module (TGSA), we have performed a comparative analysis with the initial version of \emph{TopoTxR}. This earlier version, which lacks both focal loss and TGSA, was published in a previous conference paper \citep{wang2021topotxr}. The initial \emph{TopoTxR} also employs persistent homology for extracting 1D and 2D topological cycles, as elaborated in Section \ref{sec:method:cycles}. The primary distinction lies in its approach to utilizing these masks; the initial version of \emph{TopoTxR} simply inputs the "soft" topological mask (the product of MRI images and binary topological masks) into the CNNs. In contrast, the method proposed in this paper utilizes the topological masks as an explicit guiding mechanism to train the topology-guided spatial attention module. As demonstrated in rows 10 and 13 of Table \ref{table:mainres}, our proposed method, which incorporates both the focal loss during training and the TGSA module, surpasses the performance of the initial \emph{TopoTxR} and its variants.

\begin{figure*}[!t]
\centering
\includegraphics[width=.85\textwidth]{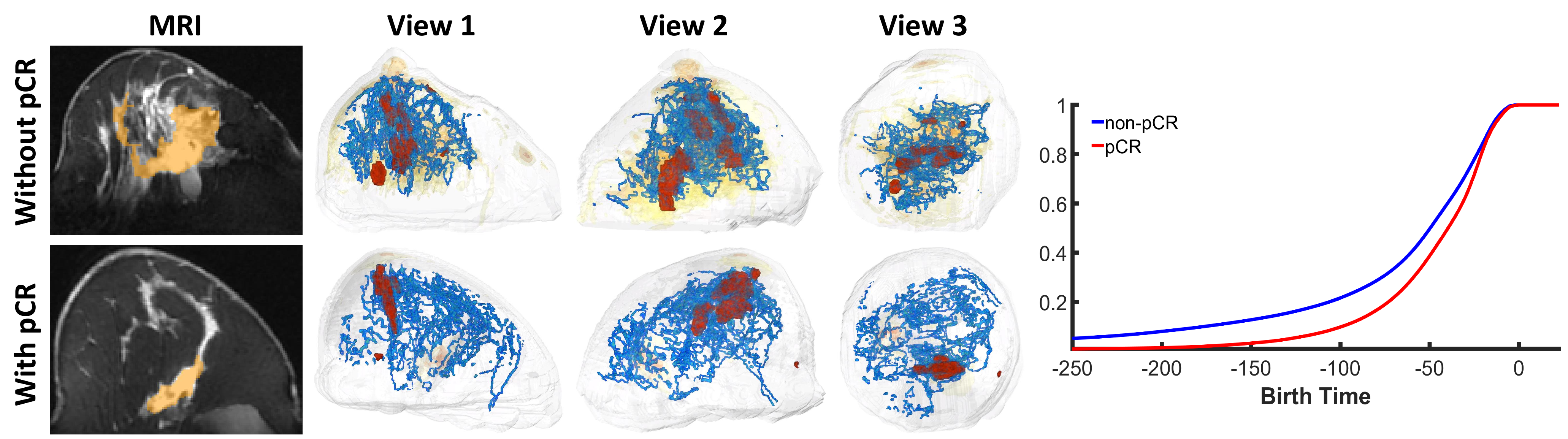}
\caption[vis-struct-new]{Qualitative comparison of patients with and without pCR. First column: Slices of breast DCE-MRIs with tumor masked in orange (tumor masks are not used in \emph{TopoTxR}). Columns 2-4: 3D renderings of topological structures from three different views. 1-D structures (loops) are rendered in blue and 2-D structures (bubbles) in red. Right: cumulative density function of topological structures' birth times.}
\label{fig:qualitative}
\end{figure*}

Observations from Table \ref{table:mainres} reveal that even in the absence of focal loss or TGSA, the initial \emph{TopoTxR} markedly outperforms the baseline methods, specifically those utilizing Radiomics, PD features, or both. Furthermore, when compared to DenseNet, a deep learning method devoid of topology information, the initial \emph{TopoTxR} still shows superior performance. These results underscore the significant value of the proposed usage of topological masks in pCR classification, which is not captured by radiomics, simple topological features like PD, or conventional CNN application.

\myparagraph{TopoTxR w/o focal loss.} We carry out an additional experiment on I-SPY1 dataset to assess the significance of the TGSA module. As indicated in rows 11 and 13 of Table \ref{table:mainres}, TGSA effectively directs the attention of the CNNs towards biologically relevant regions for pCR prediction, resulting in a substantial enhancement of performance. However, this approach still falls short of the performance achieved by the model that incorporates focal loss.

\myparagraph{TopoTxR w/o TGSA.} The performance of \emph{TopoTxR} when using only focal loss has been assessed on the Rutgers dataset, with the results detailed in row 6 of Table \ref{table:privateres}. It is evident from these results that the application of focal loss contributes to an improvement in performance.

\myparagraph{Persistence Threshold.} Recall that the persistence of a topological structure is the difference between its birth and death times. To ensure accuracy, we eliminate topological structures with low persistence, as these are typically noise-induced and could adversely affect our results. We assess the effects of persistence on the I-SPY1 dataset by applying three distinct thresholds, retaining $90\%$, $60\%$, and $30\%$ of the structures respectively. These results are detailed in Table \ref{table:abl}. As indicated in the table, retaining $30\%$ of the topological structures, which is our default setting as shown in row 13 of Table \ref{table:mainres}, offers an optimal equilibrium between the quantity and quality of these structures.

\myparagraph{Dimension of Topological Structures.} We also evaluate our method using only 1D structures (loops) and only 2D structures (bubbles). Each of these approaches outperform the baseline methods that do not incorporate topology information, as evidenced in row 3 and 4 of Table \ref{table:abl}. However, they are still not as effective as \emph{TopoTxR}, which utilizes both 1D and 2D topological structures. This outcome indicates that 1D and 2D structures provide complementary information, which is crucial for accurate pCR prediction. {For comparison, we extract 1D and 2D soft topological masks as approximations of fibroglandular tissues and stack them alongside the MRI volumes to form three-channel inputs for training a 3D CNN. The results are documented in row 5 of Table~\ref{table:abl}. While this method shows notable improvement over using MRIs alone (as shown in row 6), due to the additional information provided by the fibroglandular tissue approximation, it does not direct the model's attention as intentionally as the proposed model, resulting in inferior outcomes.}

\myparagraph{Dilation Radius.} In the process of generating topological masks, rather than using binary masks directly, we create 'soft' masks by multiplying these binary masks with the corresponding MRIs. We conduct an ablation study using I-SPY1 dataset to examine the impact of varying the radius of dilation operation on binary masks prior to this multiplication. The data presented in row 7, 8, and 9 of Table \ref{table:abl} indicates that the best performance is achieved when no dilation operation is applied to the binary masks.

\myparagraph{Insertion Point for Topology-Guided Spatial Attention Module.} Finally, we examine how the insertion point of the TGSA module after different layers affects performance. In addition to our default setting, where TGSA receives feature maps from the third convolution layer, we also examine its placement after the second and fourth layers, adjusting the size of the topological masks accordingly. The findings of this study are detailed in the last two rows of Table~\ref{table:abl}. Our observations indicate that the default setting, as shown in row 13 of Table \ref{table:mainres}, yields significantly better results.

In summary, the most effective results are obtained when 30\% of the topological structures are retained, by employing a combination of both 1D and 2D structures without any dilation, and by applying the topology-guided spatial attention module after the third convolution layer.


\section{Qualitative Analysis}
\label{sec:discussion}

\begin{figure*}[!t]
\centering
\includegraphics[width=0.99\textwidth]{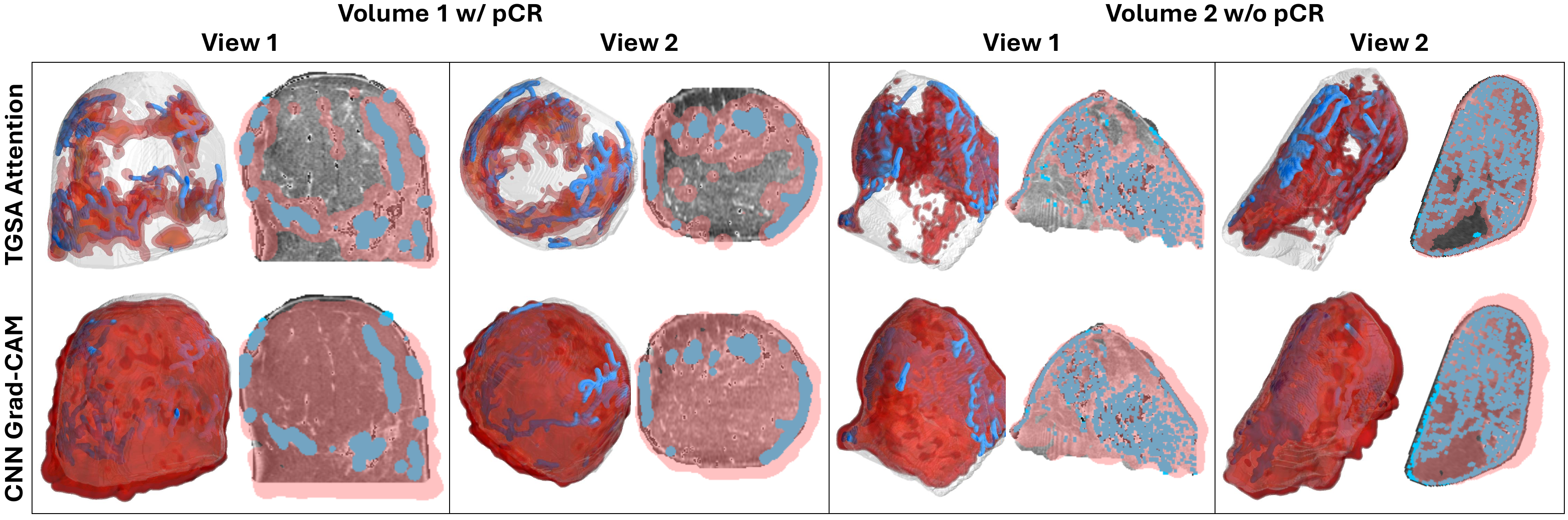}
\caption[ph]{{Attention maps of CNNs when making decisions about pCR predictions. Red: voxels contributing to decisions of CNNs; blue: voxels corresponding to extracted topological structures. Columns 1, 3, 5, and 7 display 3D renderings of the attention maps, while columns 2, 4, 6, and 8 present cross-sections corresponding to the 3D renderings on their left.}}
\label{fig:att01}
\end{figure*}

\subsection{Topological Feature Interpretation}
The topological structures identified through persistent homology effectively capture the breast tissue structures. Learning directly from these structures and their vicinity provides the opportunity for interpreting the learning outcomes and drawing novel biological insights. 
Here we provide some visual analysis as a proof of concept.

Fig. \ref{fig:qualitative} presents topological structures from three distinct perspectives, showcasing one representative sample each for cases with and without pCR. We observe that the 1D and 2D structures are sparser in the case with a pCR response, while they appear denser in the non-pCR case. Examining the corresponding MRI images, we notice that the breast with pCR exhibits scattered fibroglandular tissue and minimal background parenchymal enhancement. In contrast, the non-pCR breast displays more heterogeneous fibroglandular tissue accompanied by moderate background parenchymal enhancement. This observation suggests that the topological structures may be capturing the intricate fibroglandular structure, potentially serving as an indicator of treatment response.

We also analyze the topological characteristics of two patient populations by comparing their birth times, which indicate the threshold at which a cycle first appears. Given our use of the inverse image in the experiments, denoted as \(f=-I\), meaning the birth time essentially corresponds to $-1$ times the brightness of a structure. For each population (pCR and non-pCR),  we compute the Cumulative Density Function (CDF) curve using the average birth time from all samples in that group. As depicted in Fig.~\ref{fig:qualitative} (right), the CDF for pCR patients (in red) compared to non-pCR patients (in blue) reveals that topological structures in pCR patients’ tissues tend to be less bright or distinguishable than those in non-pCR patients, an observation that aligns with our qualitative assessments. To statistically validate these differences in CDFs, we conduct a Kolmogorov-Smirnov test \citep{massey1951kolmogorov}. The resulting $p$-value of $0.0002$ indicates a significant disparity in the birth time distributions between the pCR and non-pCR patient groups.

\subsection{Validation on the Model Attention}
The key idea in TopoTxR lies in its ability to direct CNN's attention to a much smaller set of voxels with high biological relevance manifested by the extracted topological structures. As shown in Sec.~\ref{sec:validation}, these structures have a strong connection with the breast tissue. In this section, we prove that voxels corresponding to topological structures play an essential role when CNN is making a pCR prediction. We visualize the spatial attention using a post-hoc heatmap visualization technique -- Gradient-weighted Class Activation Mapping (Grad-CAM) \citep{gradcam}. Grad-CAM generalizes Class Activation Mapping (CAM) \citep{zhoucam} by lifting its constraints on model complexity and provides good visual explanations for model decisions. It generates a coarse localization map that highlights key regions in the imaging data pivotal for predicting a specific concept. Our aim is to utilize Grad-CAM to gain interpretative insights into the decision-making process of TopoTxR. It is important to note that Grad-CAM does not alter the learning mechanism of the model. It is applied post-training to a model with fixed weights, serving purely as a tool to shed light on how the model arrives at its decisions.

We employ the Grad-CAM implementation from the M3d-CAM library \citep{gotkowski2021m3d} to visualize attention maps following the Topology-Guided Spatial Attention (TGSA) module in the best-performing \emph{TopoTxR} model. For comparison, we modify the \emph{TopoTxR} model to exclude the TGSA module, utilizing a single 3D CNN branch that processes only raw MRI data. In this setup, Grad-CAM is applied after the third convolution layer to visualize attention maps, as depicted in the second row—'CNN Grad-CAM' in Fig.~\ref{fig:att01}. Conversely, the attention maps from our proposed model are shown in the first row—'TGSA Attention.' {In Fig.~\ref{fig:att01}, we provide a comparative analysis of the attention maps from two views, showcasing one representative sample each for cases with and without pCR. This comparison highlights that the TGSA module effectively directs the CNN's focus to specific, biologically relevant regions, as delineated by topological structures. Conversely, the model lacking topology information shows a more scattered and dispersed attention distribution across the MRI volume. It is interesting to note that the topological structures appear sparser in cases with a pCR, whereas they are denser in cases without pCR.}

\section{Discussion}
Breast composition in MRI is characterized by the amount of fibroglandular tissue (FGT) and the extent of Background Parenchymal Enhancement (BPE) following contrast administration, in accordance with the American College of Radiology (ACR) Breast Imaging Reporting and Data System (BI-RADS) standards. FGT is classified into four categories: fatty fibroglandular tissue (\(\leq 25\%\)), scattered fibroglandular tissue (25-50\%), heterogeneous fibroglandular tissue (50-75\%), and extreme fibroglandular tissue (\(\geq 75\%\)). BPE is rated as minimal, mild, moderate, or marked, based on the FGT enhancement relative to the total FGT volume during the initial phase of T1-weighted fat-suppressed post-contrast imaging , which is captured 90 seconds after contrast agent administration.

Recent studies indicate that elevations in BPE correlate with an increased risk of breast cancer development and changes in BPE can predict NAC response \citep{chen2015background, liao2020background, oh2018relationship, preibsch2016background, van2015association, you2017association, you2018decreased}. Our work hinges on the idea that breast composition elucidated on MRI can predict treatment response in pre-neoadjuvant chemotherapy MRI scans. Two notable studies \citep{you2017association,you2018decreased} highlight that post-chemotherapy reductions in BPE significantly correlate with pCR, both in HER2-positive \citep{you2018decreased} and HER2-negative breast cancers \citep{you2017association}. The prevailing hypothesis is that tumor-adjacent vascular permeability might echo BPE fluctuations before and after NAC administration, with a general decline in BPE after NAC \citep{rella2020association}.

FGT resembles the concept of mammographic density, a recognized independent breast cancer risk factor \citep{duffy2018mammographic}. When assessing both FGT and BPE using MRI, research suggests that elevations in BPE, regardless of FGT presence (e.g., moderate to marked BPE in breasts that are primarily fatty or have scattered FGT), can indicate an increased cancer risk \citep{arasu2019population}. Thus, BPE is considered more instrumental than FGT for evaluating NAC response and shaping prediction models \citep{arasu2019population}. 

\section{Conclusion}
This paper introduces \emph{TopoTxR}, a novel topological biomarker that capitalizes on the rich geometric information inherent in structural MRI to enhance downstream CNN processing. To harness the intrinsic topological information effectively, we integrate a topology-guided spatial attention mechanism. Our model combines information from raw MRIs and topological masks, addressing the sample imbalance problem in datasets using focal loss. Specifically, we compute 1D cycles and 2D bubbles from breast DCE-MRIs employing the theory of persistent homology. These topological structures are then utilized to guide neural network attention by supervising the generation of attention maps. Furthermore, we demonstrate the predictive power of \emph{TopoTxR} in forecasting pCR using treatment-na\"ive imaging.

\section*{Acknowledgments}
This work received partial support from the following grants: National Science Foundation (NSF) 2144901 and National Cancer Institute (NCI) R21CA258493, R03CA223052, and R01CA297843. Ling was supported in part by National Science Foundation (NSF) Division of Information and Intelligent Systems (IIS) grants 2331769 and 2006665.

\bibliographystyle{model2-names.bst}
\biboptions{authoryear}
\bibliography{main, MEDIA}

\end{document}